\documentclass[runningheads]{llncs}
\newif\ifEditMode

\EditModefalse

\usepackage{preamble}
\usepackage{aliases}

\begin{document}
\title{Dissecting Bitcoin and Ethereum Transactions: On the Lack of Transaction Contention and Prioritization Transparency in Blockchains}
\titlerunning{Dissecting Bitcoin and Ethereum Transactions}
%
\author{Johnnatan Messias\inst{1}
\and Vabuk Pahari\inst{1}
\and Balakrishnan Chandrasekaran\inst{2}
\and \\ Krishna P. Gummadi\inst{1}
\and Patrick Loiseau\inst{3}}

\authorrunning{J. Messias et al.}

\institute{MPI-SWS \email{\{johnme, vpahari, gummadi\}@mpi-sws.org}
\and Vrije Universiteit Amsterdam \email{b.chandrasekaran@vu.nl}
\and Inria, FairPlay Team \email{patrick.loiseau@inria.fr}
}

\maketitle              
\begin{abstract}
  In permissionless blockchains, transaction issuers include a fee to incentivize miners to include their transactions.
To accurately estimate this prioritization fee for a transaction, transaction issuers (or blockchain participants, more generally) rely on two fundamental notions of transparency, namely contention and prioritization transparency.
Contention transparency implies that participants are aware of every pending transaction that will contend with a given transaction for inclusion.
Prioritization transparency states that the participants are aware of the transaction or prioritization fees paid by every such contending transaction.
Neither of these notions of transparency holds well today.
Private relay networks, for instance, allow users to send transactions privately to miners.
Besides, users can offer fees to miners via either direct transfers to miners' wallets or off-chain payments---neither of which are public.
In this work, we characterize the lack of contention and prioritization transparency in Bitcoin and Ethereum resulting from such practices.
We show that private relay networks are widely used and private transactions are quite prevalent.
We show that the lack of transparency facilitates miners to collude and overcharge users who may use these private relay networks despite them offering little to no guarantees on transaction prioritization.
The lack of these transparencies in blockchains has crucial implications for transaction issuers as well as the stability of blockchains. 
Finally, we make our data sets and scripts publicly available.

\end{abstract}

\section{Introduction}\label{s:intro}

The rate at which users issue transactions in permissionless blockchains, e.g., Bitcoin~\cite{Nakamoto-WhitePaper2008} and Ethereum~\cite{Wood@Ethereum}, is often much higher than the rate at which miners can include them in a block~\cite{Easley19a,Monopolywithoutmonopolist,Lavi-WWW2019,messias-sdbd-2020,Messias@IMC2021}.
Users typically issue transactions using a wallet software~\cite{Wallet@BitcoinCore,Wallet@Electrum,Wallet@Metamask,Wallet@Trezor}, whose primary functionality is determining an ``appropriate'' fee for a given transaction.
We use the term ``fee'' to refer generally to the incentive offered by a user to miners for prioritizing the inclusion of their transaction in a block, albeit its exact form may vary, e.g., \stress{fee rate} in Bitcoin and \stress{gas price} in Ethereum.
This (prioritization) fee varies, unsurprisingly, as a function of the level of congestion in the blockchain~\cite{Messias@IMC2021} as well as the distribution of fees across available transactions.
Inferring either of these is, however, deceptively complicated.

At first glance, these tasks appear straightforward, since every transaction is broadcast to all miners in the blockchain.
A user could simply gather all transactions broadcast over time and reconstruct the set of uncommitted transactions available to a miner (i.e., contents of the miner's \mpool) at any point of time~\cite{messias-sdbd-2020}.
We refer to this assumption of a public and uniform view (across miners) of all available transactions as \newterm{contention transparency}.
If contention transparency exists, a user could rank order available transactions by their fee (based on which miners should select transactions for inclusion) and estimate the commit delay of any transaction~\cite{Messias@IMC2021}.
Consequently, they could determine the fee that they must pay to guarantee inclusion of their transaction in a given block.
We label this assumption that the (prioritization) fee offered by a transaction is only that publicly declared by that transaction as \newterm{prioritization transparency}.
Neither the contention transparency nor the prioritization transparency, however, holds today in permissionless blockchains.

\parai{Lack of contention transparency.}
Not all transactions are publicly broadcast.
Users can submit transactions to a subset of miners or mining pools via \stress{private channels} or \newterm{relays} that are opaque to the public (i.e., transactions remain private to the relay, until they are committed)~\cite{EdenNetwork,EthermineMEVRelay@Ethereum,Flashbots@Ethereum,Taichi@accelerator}.
Users may also submit their transaction to a specific mining pool that assures them a fast commit time \cite{BTC@accelerator,EdenNetworkWhitepaper,F2Pool@accelerator,ViaBTC@accelerator}.
This paper reveals that such private mining practices (i.e., where transactions are submitted to only a subset of the miners) are becoming commonplace and analyzes the characteristics of these private transactions.

\parai{Lack of prioritization transparency.}
The fees offered by a transaction could be substantially more than that publicly declared by it.
A transaction could, for instance, privately offer additional fees to a miner to ``accelerate'' its inclusion in a block~\cite{AntPool@accelerator,BTC@accelerator,F2Pool@accelerator,Poolin@accelerator,ViaBTC@accelerator}.
Many such transaction-accelerator (or \newterm{front-running as a service (FRaaS)}) platforms exist for Bitcoin~\cite{BTC@accelerator,F2Pool@accelerator,Poolin@accelerator,ViaBTC@accelerator} and Ethereum~\cite{Eskandari@FC-2020,Flashbots@Ethereum,Taichi@accelerator,strehle2020exclusive}.
Furthermore, the same transaction could offer different fees to different mining pools (via their relays).
The presence of such hidden or dark-fees could fundamentally erode the reliability of any fee prediction:
Transaction issuers may end up paying substantially large fees without receiving proportional or any reduction in commit delays.
This paper characterizes the prevalence of such dark-fee transactions and analyzes the most popular private relay network available in Ethereum, Flashbots~\cite{Flashbots@Ethereum}. Furthermore, we conduct active experiments in both Bitcoin and Ethereum to validate our assumptions regarding the prioritization transparency.
In addition to showing that transaction fees may not be uniform across miners, we claim that, given the lack of contention transparency, the lack of prioritization transparency may become more widespread than it is now.

The lack of contention and prioritization transparencies stem from real, non-trivial concerns of transaction issuers.
The risk of transactions being front-run by bots~\cite{Daian@S&P20,Eskandari@FC-2020,Christof@USENIX,Weintraub@IMC2022}, for instance, creates the need for transaction privacy.
Mining pools that address this need also facilitate, unsurprisingly, off-chain payments via which transaction issuers can (privately) incentivize the miners~\cite{AntPool@accelerator,BTC@accelerator,F2Pool@accelerator,Messias@IMC2021,Poolin@accelerator,ViaBTC@accelerator}.
We view these developments as natural and logical steps in the evolution of blockchains and back our assertions with empirical observations.
We claim, therefore, in contrast to prior work~\cite{Daian@S&P20,strehle2020exclusive}, that it is only the opacity of the overall fees issued by a transaction issuer that poses a fundamental threat to the stability of blockchains:
Transaction issuers cannot, for instance, precisely infer the fee required to commit their transactions into the next block, and miners can, consequently, overcharge them as the ``real'' fees are opaque to the rest of the network~\cite{Weintraub@IMC2022}.

We summarize our contributions as follows.
We characterize the lack of contention transparency in both Bitcoin and Ethereum:
We show that the use of private channels or relay networks to submit transactions directly to a subset of miners is becoming widespread.
This practice will likely erode prioritization transparency, as transaction issuers may not be able to estimate the appropriate fees, none of which are publicly visible.
We characterize the prevalence of such private transactions fees. We found that Flashbots bundles represent \num{52.11}\% of all Ethereum blocks.
With the lack of prioritization transparency, miners might overcharge users when they send their transactions privately. We also show that Bitcoin miners collude (with an aggregate hashing power of more than \num{50}\% of the network's total hashing power) when including dark-fees transactions.
Finally, we release our data sets and the scripts used in our analysis to enable the scientific community to reproduce our results~\cite{Messias-DataSet-Code-2023}.

\section{Related Work} \label{s:related}

\begin{sloppypar}
There is a rich literature on block rewards as incentives for mining~\cite{Chen@AFT19,Eyal-CACM2018,Fiat@EC19,Goren@EC19,Kiayias@EC16,Noda@EC20,Pass_Seeman_Shelat_2017,Romiti2019ADD,sompolinsky2015secure,Zhang_Preneel_2019}.
Recent work also analyzed the implications of relying on transaction fees separately~\cite{Carlsten@CCS16} and in conjunction with block rewards~\cite{Tsabary@CCS18}, as well as the relationship between such incentives and transaction waiting times~\cite{Easley19a}.
These prior work assume that transactions are broadcast to all miners and the fees offered is uniform across miners.
None of them acknowledge the issue of transparency.
\end{sloppypar}

Basu~\ea~\cite{Basu-CoRR2019} and Lavi~\ea~\cite{Lavi-WWW2019} addressed the inefficiencies in transaction-fee setting mechanisms (i.e., first-price auctions) by proposing alternative mechanisms.
They claim that miners might be dishonest, albeit they present no empirical evidence.
Siddiqui~\ea{}~\cite{Siddiqui@AAMAS20} used simulations to show that, if transaction fees are the only incentives, miners will select transactions greedily, thereby increasing the commit times of many transactions.
Prior work also analyzed the Ethereum fee (i.e., gas price) mechanism to determine the gas price for a given transaction~\cite{Pierro@IWBOSE,Liu@DSA,Mars@COMPSAC,Turksonmez@COINS}.
The fee estimation and fee-based prioritization schemes in these studies do not take into account dark-fees or private mining.

\begin{sloppypar}
Many transaction-accelerator, or FRaaS, platforms exist for both Bitcoin~\cite{BTC@accelerator,F2Pool@accelerator,Poolin@accelerator,ViaBTC@accelerator} and Ethereum~\cite{Eskandari@FC-2020,Flashbots@Ethereum,Taichi@accelerator}.
Transaction issuers might resort to such acceleration or off-chain payment channels to hide their true fee from competitors and avoid being front-run~\cite{Daian@S&P20,strehle2020exclusive}.
Tim Roughgarden~\cite{Roughgarden@EC21} discussed the incentives for off-chain agreements (such as dark-fees) between miners and users for first-price auctions and different deviations of the new Ethereum fee mechanism \stress{EIP-1559 protocol}~\cite{EIP-1559}.\footnote{The EIP-1559 went live in the Ethereum's London hard fork upgrade on August 5\tsup{th}, 2021, at block number \href{https://etherscan.io/block/12965000}{\num{12965000}}.}
Roughgarden showed that miners and users cannot strictly increase their joint utility through off-chain payments under EIP-1559 because on-chain bids can be easily replaced by the off-chain bids. 
However, utility here is only based on the revenue of bidding for block space. The author did not take into account that utility might depend on other factors, such as transaction issuers wanting to keep their actual bids for block space hidden through off-chain payments, which strictly increases their chances of prioritization, as other bidders cannot counter bid, as they are unaware of the bid itself.
\end{sloppypar}

Closest to our work are two that analyze private mining.
Strehe and Ante~\cite{strehle2020exclusive} investigated \stress{exclusive mining} (or private mining), where transactions issuers and miners collude to include transactions that have been sent through a private network.
In this case, the transactions are not publicly disclosed until they have been included in a block; besides, the fees can remain opaque to everyone forever, as such off-chain agreements may use fiat currencies.
Weintraub~\ea~\cite{Weintraub@IMC2022} measured the popularity of \stress{Flashbots}, the most used private relay network for Ethereum.
Our work, in contrast, extensively investigates private transactions in both Bitcoin and Ethereum blockchains.
Through active measurements, we empirically show that Bitcoin miners collude and highlight the colluding mining pools.
We show that Flashbots bundles are quite prevalent in Ethereum and are mainly used for calling Decentralized Exchanges (DEX) contracts to take advantage of \stress{Maximal Extractable Value (MEV)} opportunities.
Finally, we discuss why our findings are still valid after ``The Merge''---an Ethereum hard fork deployed on September~15\tsup{th},~2022~\cite{Eth-PoS,Eth-Merge}.

\section{On contention transparency}
\label{sec:contention_transparency}

\subsection{The Rise of Private Relay Networks}

With the lucrative market of Decentralized Finance (DeFi) in Ethereum, today, bots engage in predatory front-running behaviors such as sandwich attacks and transaction-replay attacks \cite{Daian@S&P20,kiffer2017stick,Qin@BEV,Qin@FC21,Christof@USENIX,Weintraub@IMC2022,Zhou@S&P2021}. 
Relay networks help users to counter such attacks:
They provide users with a private channel for communicating with miners, who have to prove their identity to participate in the relay.
Relay networks help users completely bypass the P2P network: Users send their transactions to the relay network, which in turn relays them to its participant miners.
The relay network and its participants claim (a) not to front-run these transactions and (b) to keep them private until they are included in a block~\cite{Flashbots@API}.
These transactions, hence, by construction, experience no front-running issues.
Relay networks are centralized; if miners misbehave, they may lose their network membership and forfeit their future profits.
Multiple relay networks (e.g., bloXroute~\cite{BloXroute@Ethereum}, Taichi Network~\cite{Taichi@accelerator}, and others~\cite{EdenNetwork,EthermineMEVRelay@Ethereum}) exist today, but we focus on Flashbots~\cite{Flashbots2Docs@Ethereum}, the largest relay network for Ethereum.

\parai{Flashbots}
Flashbots's users \stress{bundle} one or more transactions in some specific order~\cite{Flashbots@Ethereum}.
Miners are expected to mine the entire bundle (retaining the ordering of transactions within the bundle) and place it at the top of their blocks.
The miners receive a fee (paid via a direct transfer to their wallets) for including the bundle in addition to the (traditional) fees associated with the transactions in that bundle.
If there are two competing bundles---capturing the same financial opportunity, e.g., liquidations---miners will choose the one with the highest reward (i.e., maximizing financial incentives).
The other bundle is \stress{discarded} (since the financial opportunity no longer exists after having been captured by the included bundle), albeit its transactions do \stress{not} expend \stress{any} gas.
Therefore, except for a network base fee introduced in EIP-1559, arbitrageurs and liquidators can participate without having any balance in their wallet:
If they successfully capture a financial opportunity, they pay the miner from the profit secured and pocket the rest~\cite{Flashbots2Docs@Ethereum}.
Flashbots is a \stress{free} to use relay network, and they allow anyone to query whether a transaction used their relay network and the private fees paid to the miner (after it has been committed in a block).
We use this publicly available data for analyzing the transactions issued (privately) on Flashbots.
Flashbots, however, does not list the discarded bundles (or its transactions): we have access, hence, only to committed transactions.

\subsection{Characterizing Private Relay Networks}
\label{ss:characterize-pvt-relay}

\begin{figure*}[tb]
  \centering
  \begin{subfigure}[b]{\onecolgrid}
    \includegraphics[width={\textwidth}]{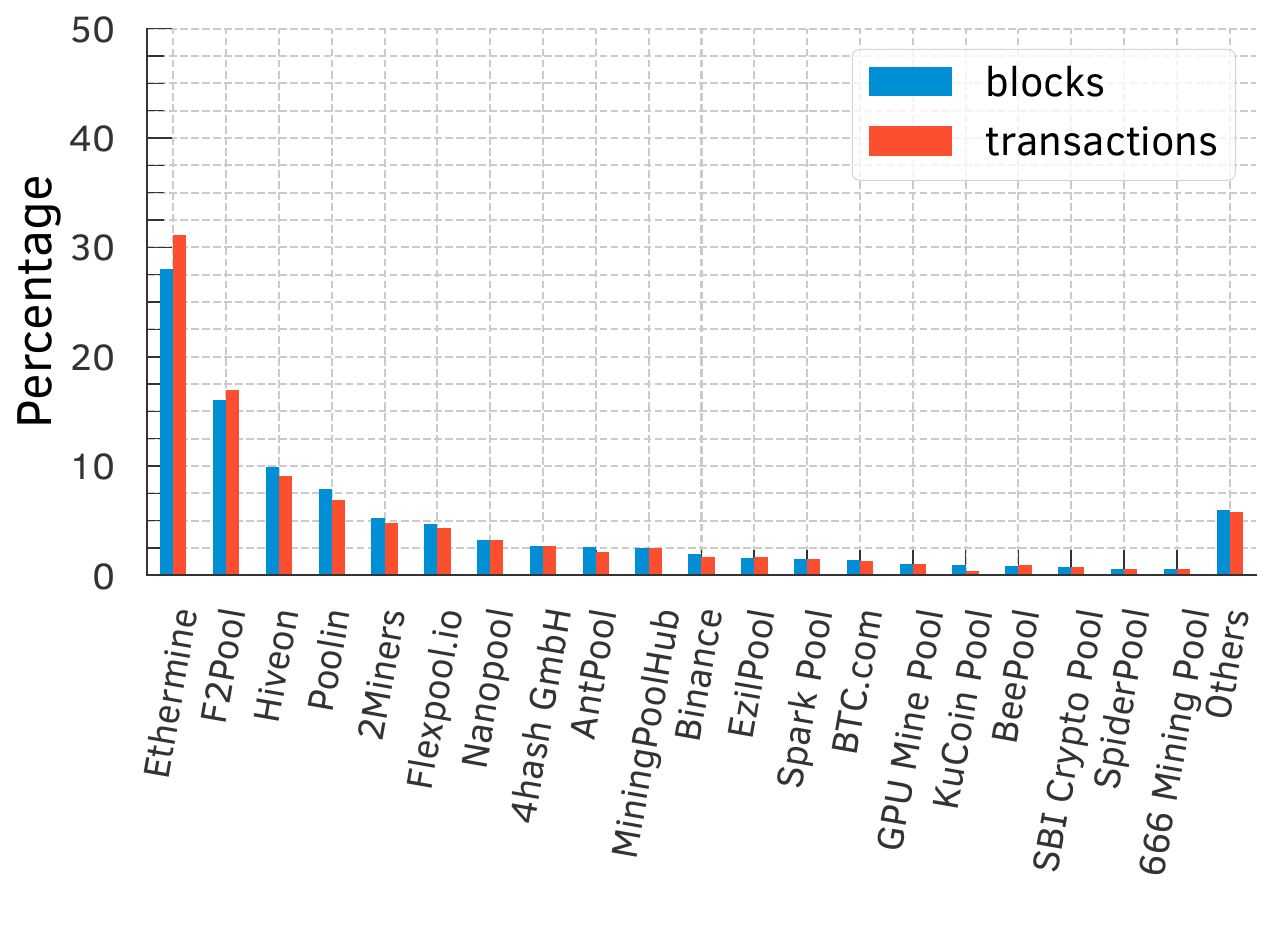}
    \figcap{Ethereum distribution}\label{fig:dist-tx-blks-ethereum}
  \end{subfigure}
  \begin{subfigure}[b]{\onecolgrid}
    \includegraphics[width={\textwidth}]{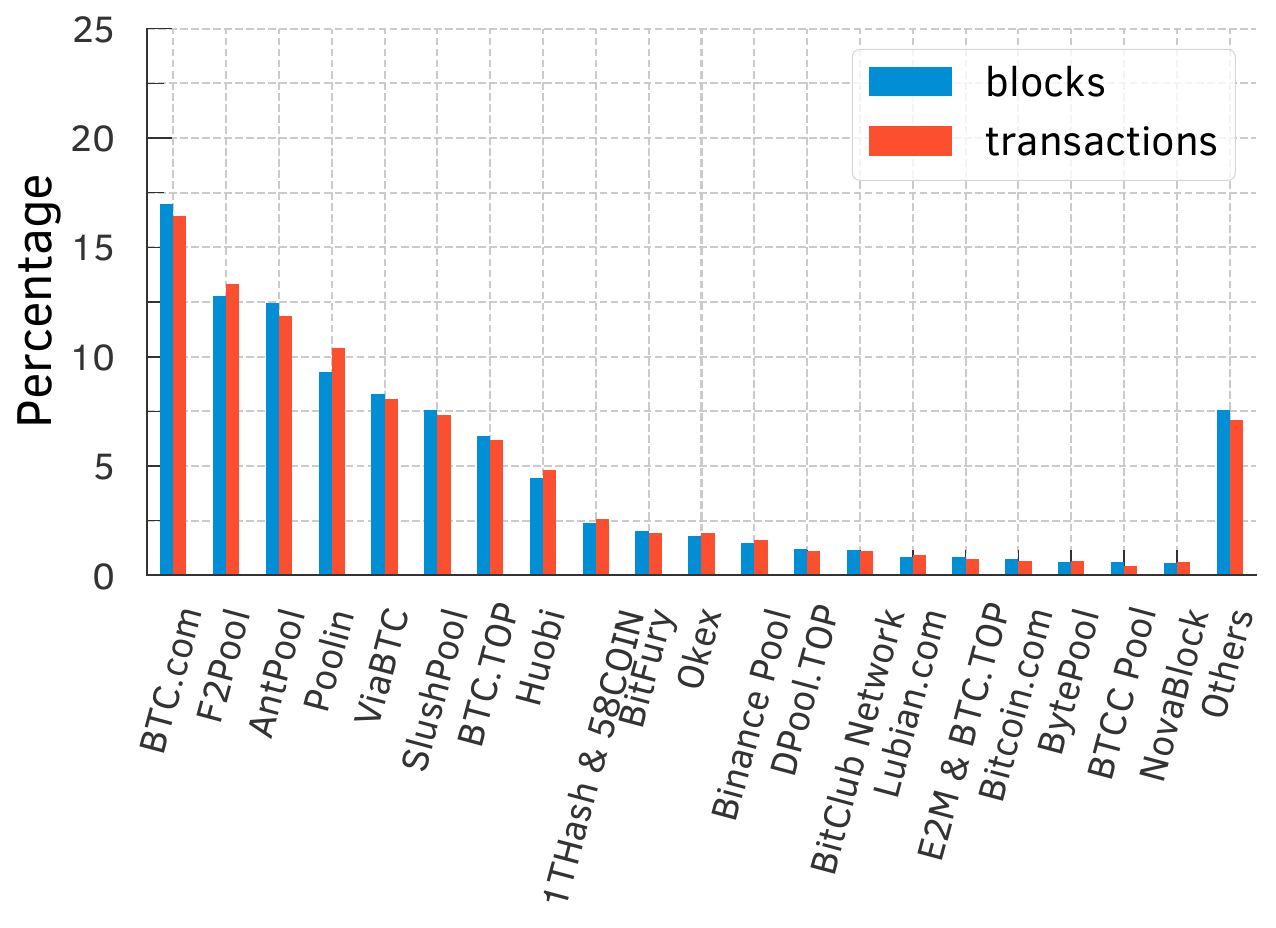}
    \figcap{Bitcoin distribution}\label{fig:dist-tx-blks-bitcoin}
  \end{subfigure}
  \figcap{Blocks mined and transactions confirmed in (a) Ethereum and (b) Bitcoin by the top-20 mining pools; ``Others'' consolidates the remaining mining pools.}
  \label{fig:eth-btc-blocks-txs-distribution}
\end{figure*}

We gathered all Ethereum blocks mined over a $9$-month time period---from September $8\tsup{th}$, 2021 to June $30\tsup{th}$, 2022---to investigate the behavior of Ethereum mining pools.
This data set contains \num{347629393} issued transactions and \num{1867000} blocks (from block number \num{13183000} to \num{15049999}).
We used miners' wallet addresses to infer the block owners, but we failed to identify the owners of \num{46895} blocks (or \num{2.51}\% of the total); we grouped the latter into one category, ``Unknown.''
Figure~\ref{fig:dist-tx-blks-ethereum} shows the distribution of blocks and transactions mined in Ethereum by the top-20 mining pools.
We also retrieved \num{6937292} transactions ($2\%$ of all issued transactions) contained in \num{3284886} bundles from Flashbots; these are transactions sent privately to miners.
\num{972911} ($52.11\%$) of blocks in the data set have at least one such Flashbots transaction:
\stress{Private transactions are becoming quite common across most of the powerful mining pools in Ethereum.}

Flashbots labels its bundles (and constituent transactions) into one of three categories: (i) \stress{flashbots}, which represent those sent through their private relay;  (ii) \stress{rogue}, referring to those delivered to a (Flashbots) miner, but via a different relay network; and (iii) \stress{miner payout}, indicating a bundle containing payouts to users of a mining pool~\cite{Weintraub@IMC2022}.
We find \num{58.82}\%, \num{27.93}\%, and \num{13.25}\% of transactions belonging to the flashbots, miner payout, and rogue categories, respectively.
We also noticed that \num{70260} (\num{1.01}\%) of all Flashbots transactions failed to execute after inclusion in a block.
A small fraction of transactions is, hence, not successfully executed despite using private relays.

Flashbots claims to have $\approx85\%$ of the total Ethereum hash rate~\cite{Flashbots2Docs@Ethereum}.
Per our analyses, however, the majority of the mining pools ($47$ out of $48$---barring EthPool) use Flashbots, accounting for $99.99\%$ of the total Ethereum hash rate,
A recent work also corroborates our findings~\cite{Weintraub@IMC2022}.

Some of the most powerful mining pools like Spark Pool\footnote{Spark Pool suspended their mining services on Sept. 30\tsup{th}, 2021, due to regulatory requirements introduced by Chinese authorities~\cite{SparkPool@CoinTelegraph}.} (which cooperates with Taichi Network~\cite{Taichi@accelerator}), Ethermine~\cite{EthermineMEVRelay@Ethereum}, and F2Pool (part of Eden Network~\cite{EdenNetwork}) offer their own relay networks. 
As these networks allow transaction issuers to send transactions exclusively to a specific miner, we hypothesize that miners would prefer (or prioritize) these transactions to those sent via the public P2P network. 
Crucially, payments from these private transactions are guaranteed, while those from publicly issued transactions are not---they are available to any miner willing to commit them.
\stress{Miners, hence, would likely offer preferential treatment for private transactions.}

\subsection{On preferential treatment of private transactions} \label{subsec:preferential_treatment}

We substantiate our hypothesis of preferential treatment for private transactions via an active experiment conducted on September 8\tsup{th}, 2021.
We issued $8$ transactions, where $4$ were sent privately via the Taichi Network, powered by Spark Pool, and $4$ through the public Ethereum network (refer Table~\ref{tab:acceleration-experiment-ETH} in~\S\ref{sec:private-txs}).

While running the experiment, we checked if the popular Ethereum blockchain explorers (i.e., Etherscan~\cite{Etherscan@ETH-explorer}, Blockchain.com~\cite{Blockchain@ETH-explorer}, and Blockchair~\cite{Blockchair@ETH-explorer}) observed any of our private transactions;
if they did, it would imply that the Taichi Network leaked the transactions to the public.
While the public transactions appeared in these blockchain explorers, right after we sent them through the public P2P network, the private transactions were not observed by any of them until the transactions were included in a block.
More importantly, our private transactions were \stress{not} flagged by Etherscan (which relies on Flashbots API~\cite{Flashbots@API} and more recently on EigenPhi~\cite{EigenPhi@Ethereum}) as private, \stress{even after inclusion in a block}.
Measuring the prevalence of private transactions is, hence, challenging; it is likely that our estimates of the volume of private transactions based on such tools represent, hence, a lower bound.

Babel Pool included \num{2} out of our \num{4} private transactions.
Spark Pool technically supports this mining pool, implying that they ``collaborate'' in committing private transactions sent over the Taichi network~\cite{Babel@Pool}.
Our transactions were included, however, in the appropriate position in the block based on their fees.
We delve into the prioritization of transactions in the next section.

We also characterize the prevalence of private transactions in Ethereum and indicate that mining pools can each have a distinct set of private transactions in their \mpool.
Users, as a result, can no longer rely on the public \mpool alone to estimate their transaction fee.
Given the absence of other data, they are highly likely to end up with a false estimate of the ``appropriate'' transaction fees for their transactions.

\begin{figure*}[tb]
  \centering
  \begin{subfigure}[b]{\onecolgrid}
    \includegraphics[width={\textwidth}]{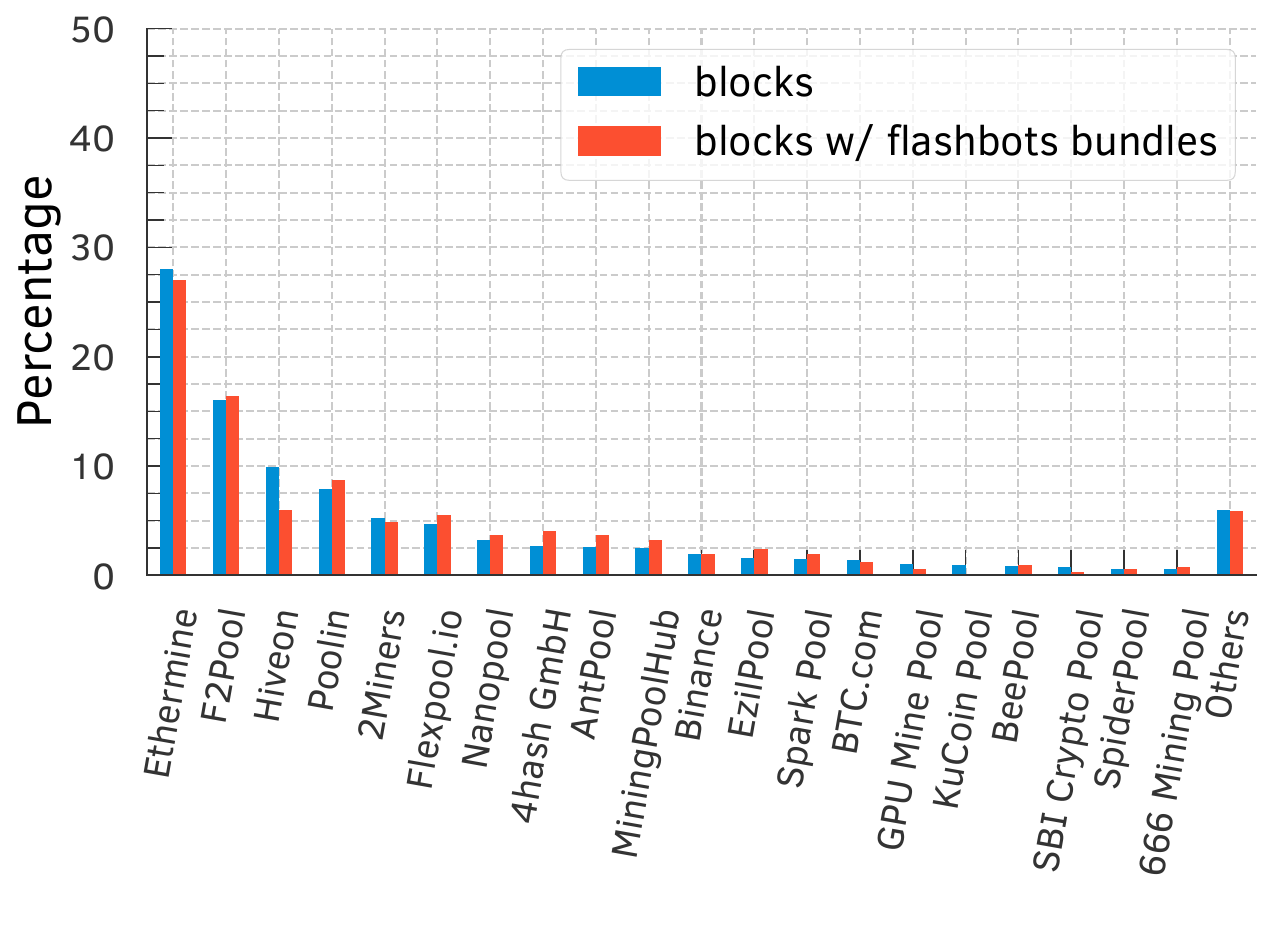}
    \figcap{Distribution of Flashbots blocks}\label{fig:dist-flashbots-blocks-ethereum}
  \end{subfigure}
  \begin{subfigure}[b]{\onecolgrid}
    \includegraphics[width={\textwidth}]{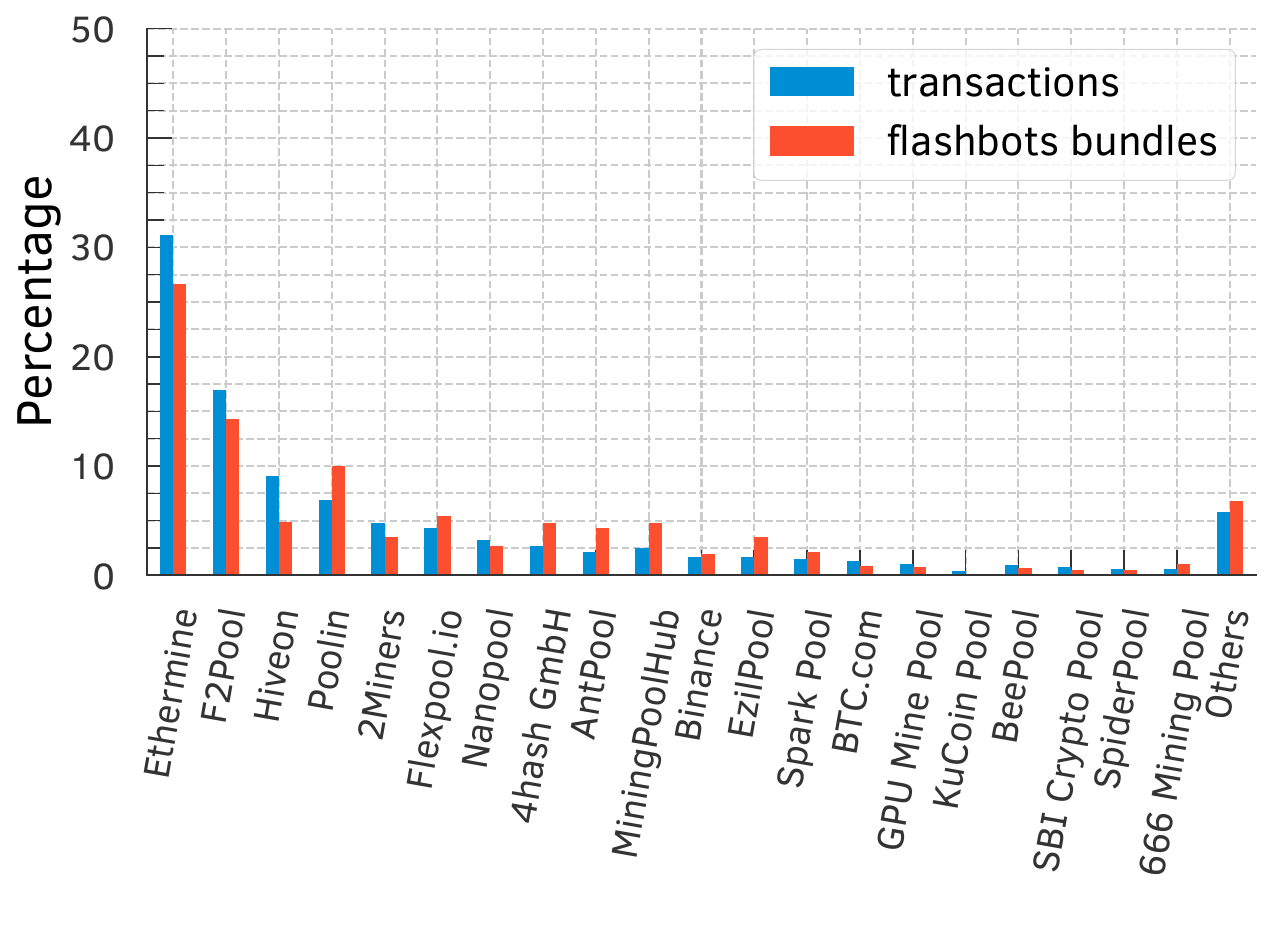}
    \figcap{Distribution of Flashbots bundles}\label{fig:dist-flashbots-bundle-ethereum}
  \end{subfigure}
  \figcap{
  Distribution of (a) blocks with at least one Flashbots bundle and (b) bundle of transactions per block, per mining pool. Ethermine included 27.05\% of all blocks with a Flashbot bundle and 26.63\% of all Flashbots bundles, while mining around 28.05\% and 31.11\% of all blocks and transactions, respectively.}
  \label{fig:dist-flashbots-txs-bundle-ethereum}
\end{figure*}

\section{On prioritization transparency}
\label{sec:prioritization_transparency}

\subsection{Prevalence of transaction bundling}\label{subsec:prevalence_of_flashbots}

Flashbots bundles are prevalent in Ethereum (refer~\S\ref{ss:characterize-pvt-relay}).
Each Flashbots bundle contains at least $1$ transaction and at most \num{631} transactions; on average they contain \num{2.11} transactions, with a median of \num{1} and a standard deviation of $6.47$.
We noticed that Ethermine alone included more than a quarter ($26.63\%$) of all \num{3284886} bundles (Figure~\ref{fig:dist-flashbots-txs-bundle-ethereum}).
Also, blocks contain at most $40$ bundles, with an average of \num{3.38}, a median of $3$, and a standard deviation of \num{2.64}.

\paraib{Maximal Extractable Value (MEV)}
Flashbots allows users to bundle together a set of transactions, thereby specifying the order in which they are executed.
The bundles can also include public transactions,  propagated over the public P2P network.
A public transaction that buys a coin on a DEX can, for example, lead to an arbitrage opportunity~\cite{Qin@FC21}.
A user can include this transaction in a bundle along with one of their own to capture this arbitrage opportunity.
The last transaction in the bundle usually pays the miner (based on the profit made) in ether via a direct transfer (i.e., \stress{coinbase transfer})  to their wallet addresses~\cite{Flashbots2Docs@Ethereum}.
This essentially means that miners are being offered different prices for mining the same transaction.
In other words, miners have a financial incentive for including transactions that are in a bundle at the top of a block, even though the public fee offered through gas price in the transaction data is very low. 
Hence, each transaction in the bundle has a normal gas price and a \stress{bundle gas price}, which is calculated using the total gas used by all transactions in the bundle and the total miner reward for mining the bundle.

\paraib{Bundling public transactions}\label{subsec:bundling_public_txs}
To identify bundles with transactions that were probably sent through the public P2P network, we rely on a simple heuristic.
Specifically, we focus on transaction bundles of size \num{2} and \num{3}, and search for transactions that have likely resulted in a publicly sent transaction being bundled.
Then, we find bundles issued from different issuers that include a zero and non-zero \stress{max-priority fee}\footnote{The \stress{max-priority fee} was introduced in EIP-1559 as the unique financial incentive miners get for including publicly announced transactions.
The other fees are burned.} transactions.
The intuition is that miners have no incentive to include transactions that offer a zero max-priority fee, as they receive no rewards for mining these transactions.
Unless they receive extra payment (through Flashbots coinbase transfer). Hence, transactions that have a non-zero max-priority fee were likely sent publicly.

For transaction bundles of size \num{2}, we look for transactions whose issuers are not the same. Furthermore, we look for cases where the first transaction offers a non-zero max-priority fee, with no coinbase transfer to the miner, and the second transaction offers a \num{0} max-priority fee and a non-zero coinbase transfer.

For transaction bundles of size \num{3}, we look for signs of sandwich attacks~\cite{Qin@BEV}. We look for bundles where the first and last transactions have the same issuer, but the second transaction has a different issuer.
Additionally, we check that the first and third transactions offer a \num{0} max-priority fee, meaning that the miner receives no reward from the gas price for mining these transactions.
Then, we ensure that the second transaction offers miners a non-zero max-priority fee, while the third offers miners a fee through direct coinbase transfer. 
This scenario might be a classic sandwich attack, where public transactions are bundled between two private transactions, sent by the same issuer, and the miner gets paid via a coinbase transfer from the third transaction~\cite{Qin@BEV}.

\begin{figure*}[tb]
  \centering
    \begin{subfigure}[b]{\onecolgrid}
    \includegraphics[width={\textwidth}]{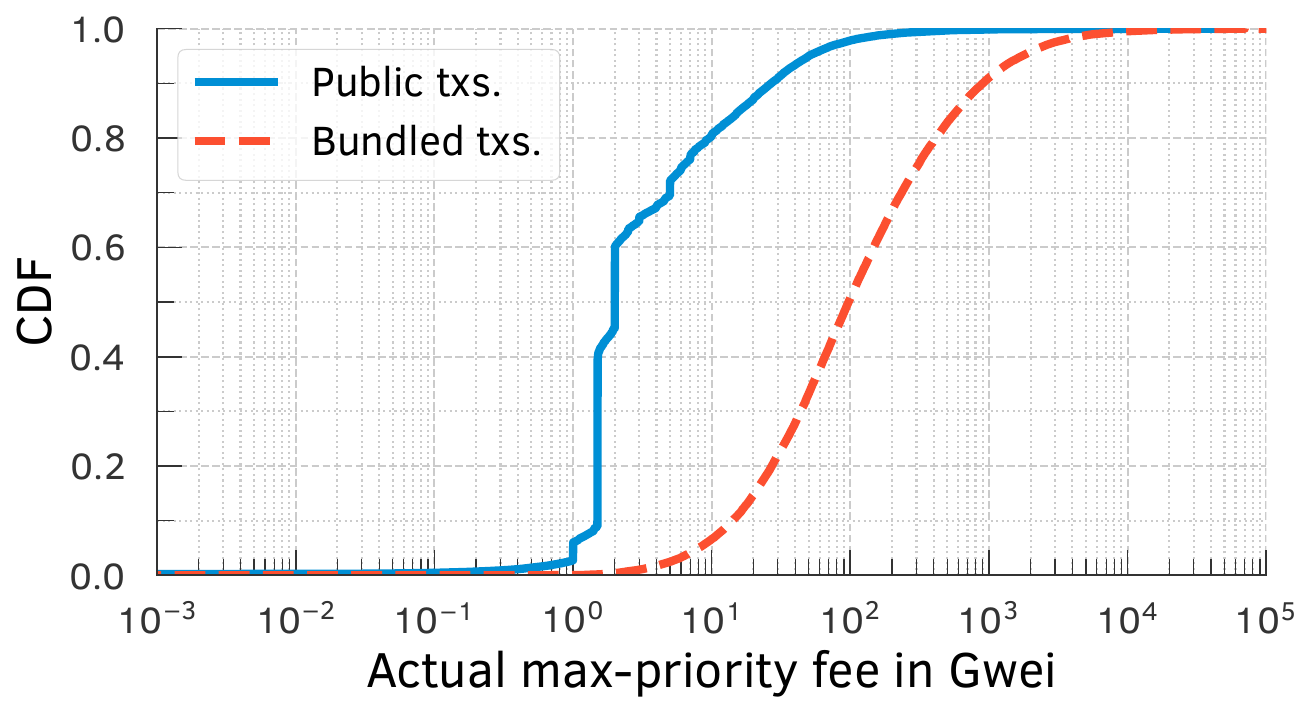}
    \figcap{Public vs bundled's actual fee
    }\label{fig:public_vs_bundle}
  \end{subfigure}
  \begin{subfigure}[b]{\onecolgrid}
    \includegraphics[width={\textwidth}]{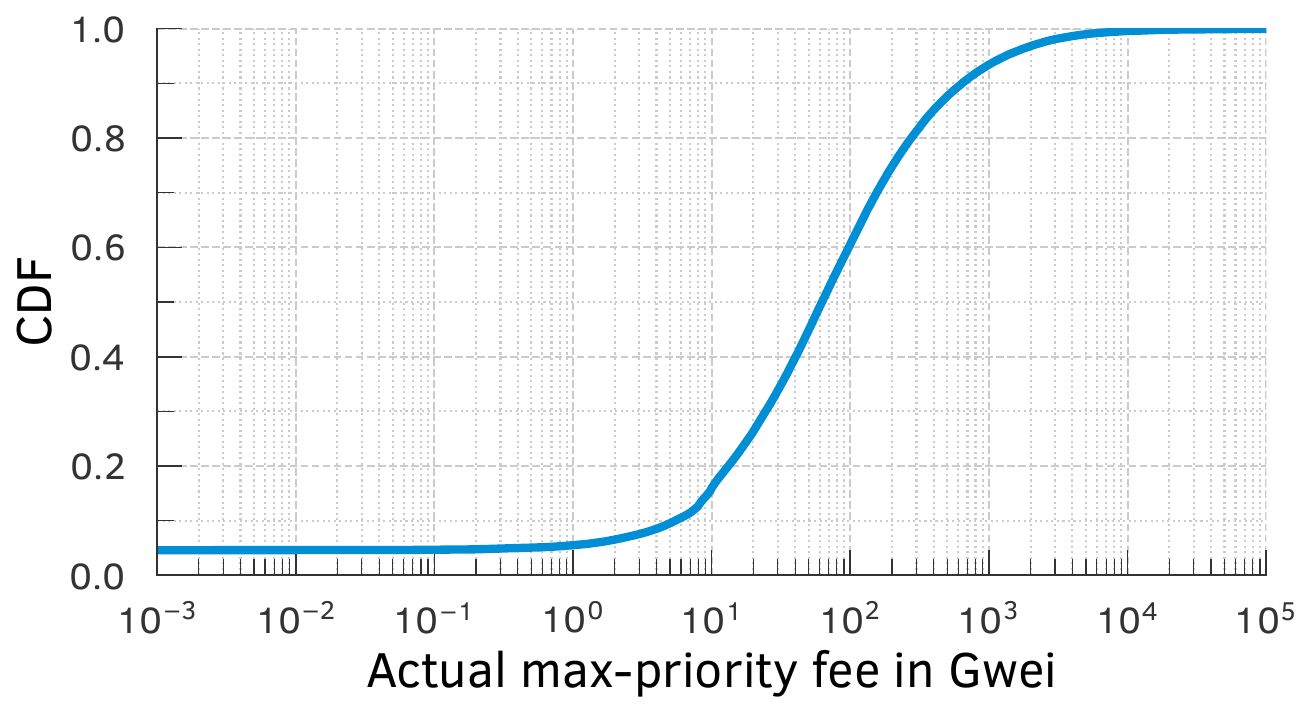}
    \figcap{Difference in actual max-priority fee}\label{fig:gas_price_diff}
  \end{subfigure}
   \figcap{Diff. between the actual max-priority fee of public transactions and Flashbots bundles; bundles typically offer a larger \stress{effective} fee to the miners.}
  \label{fig:gas_price_differences}
\end{figure*}

We found \num{853394} transactions in \num{426697} bundles of length \num{2}, and \num{1231695} transactions in \num{410565} bundles of length \num{3}. 
From those, we found that \num{110401} (\num{25.87}\%) and \num{37447} (\num{9.12}\%) bundles, of lengths \num{2} and \num{3}, respectively, fit our heuristic. 
We then calculate the \stress{actual max-priority fee} for these bundles, as the total gas used by all transactions in the bundle divided by the total miner reward (from gas usage and coinbase transfer). 
Figure~\ref{fig:gas_price_differences} shows the price difference miners get for including publicly and bundled transactions. 
Note that around \num{40}\% of transactions differ in the actual max-priority fee by \num{100} gwei-per-units-of-gas. Flashbots bundles offers much higher gas prices in comparison to the public announced max-priority fee alone.

\paraib{Towards liquidations through bundling}
Lending protocols rely on \stress{over-collateralization} of assets:
In order to borrow assets from these protocols, a user has to deposit a collateral of at least \num{150}\% of the borrowed amount.
To borrow \num{1} USDC on AAVE, for example, a user would have to collateralize at least \num{1.5} USDC worth of another asset (e.g., in ETH or BTC).
If the ratio of the collateral asset versus the borrowed asset falls below \num{1.5}, the user's position can be liquidated by any other participant until the ratio stabilizes to \num{1.5} again.
The liquidator then pays back a portion of the user's debt to receive the collateral asset at a discount.
In order to assess an asset's on-chain value, lending protocols rely on oracle services, e.g., Chainlink Data Feeds~\cite{breidenbach2021chainlink,ChainlinkDataFeeds}.
In the case of the two largest lending platforms, AAVE V2~\cite{AAVE} and Compound~\cite{Compound}, for instance, Chainlink provides the price of each asset in ETH and USD, respectively. 

We found \num{16418} liquidations in AAVE and \num{6387} liquidations in Compound.
Out of these, there were \num{4863} AAVE liquidations and \num{2036} Compound liquidations that were sent privately through Flashbots. 
In AAVE, the three largest collateral assets that were liquidated were WETH (\num{57.58}\%), LINK (\num{11.84}\%), and WBTC (\num{8.99}\%). 
The debt assets paid for, i.e., the assets borrowed by the users, were USDC (\num{33.77}\%), USDT (\num{22.27}\%), DAI (\num{19.39}\%), and GUSD (\num{5.12}\%), all of which are stablecoins and account for over \num{80}\% of the assets repaid by liquidators.
In Compound, the three largest collateral assets that were liquidated were WETH (\num{69.7}\%), WBTC (\num{10.31}\%), and UNI (\num{5.5}\%).
The debt assets were USDC (\num{38.9}\%), DAI (\num{30.45}\%), USDT (\num{23.38}\%), and TUSD (\num{2.7}\%), all of which are stablecoins and account for over \num{90}\% of the assets repaid by liquidators.

\parai{Liquidation with bundled oracle updates}
To check the adverse effect of bundling oracle updates, we looked at bundles with Chainlink oracle updates as they are a key part of liquidations.
We identified \num{1165} AAVE liquidations distributed within \num{1154} bundles (\num{2662} transactions including \num{1301} oracle updates) that contained at least one oracle update.
In Compound, we found \num{648} liquidations distributed within \num{641} bundles (\num{1457} transactions including \num{751} oracle updates) that contained oracle updates.
In AAVE, out of \num{1154} bundles, there were \num{994} (86.14\%) bundles that contained an oracle update followed by a liquidation, and \num{52} (4.51\%) with two oracle updates followed by liquidations.
In Compound, out of \num{641} bundles, there were \num{548} (85.49\%) bundles that contained an oracle update followed by a liquidation, and \num{39} (6.08\%) with two oracle updates followed by liquidations.
For details on the specific liquidations for both AAVE and Compound, please refer~\S\ref{sec:liquidations-cll-updates} in the appendix.
Out of the total \num{1813} liquidations in AAVE and Compound we found that only \num{24} were possible in the previous block.
Almost \num{98.68}\% of such liquidations were, hence, only possible because of the Chainlink updates in that block.

\begin{figure*}[tb]
  \centering
  \begin{subfigure}[b]{\onecolgrid}
    \includegraphics[width={\textwidth}]{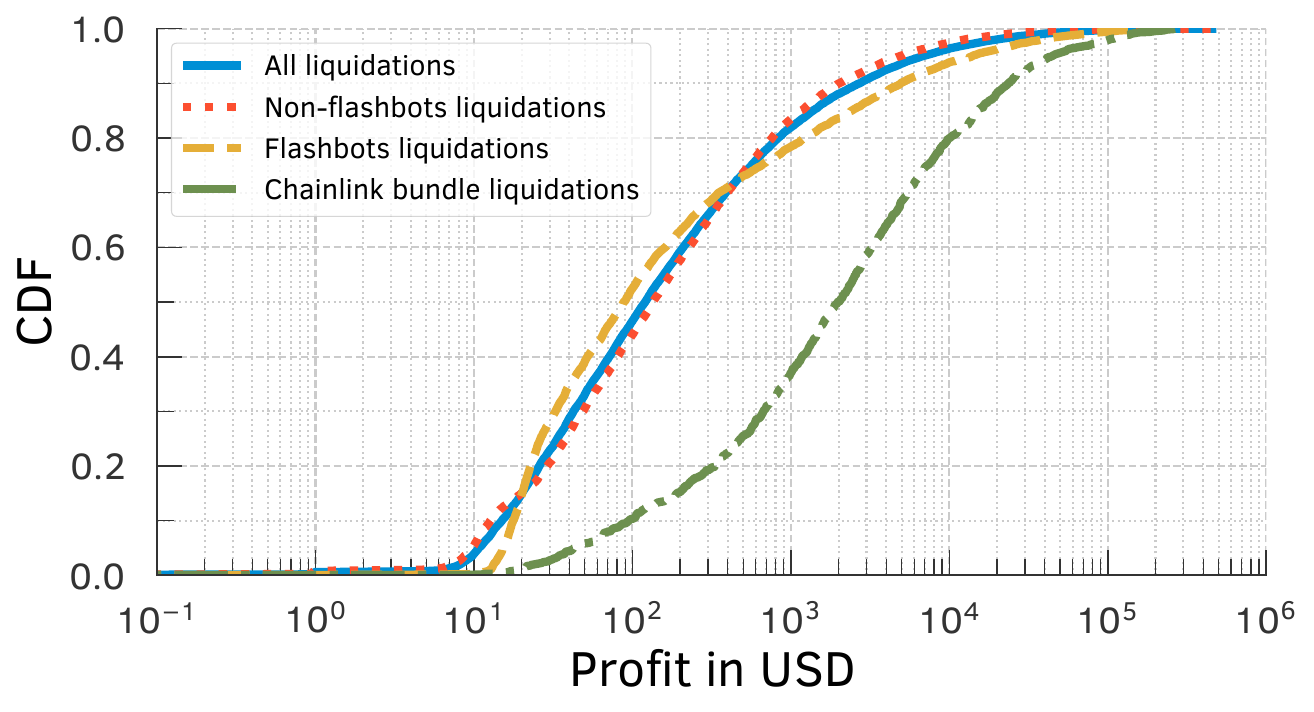}
    \sfigcap{Liquidations profit in AAVE}\label{fig:oracle-update-AAVE}
  \end{subfigure}
  \begin{subfigure}[b]{\onecolgrid}
    \includegraphics[width={\textwidth}]{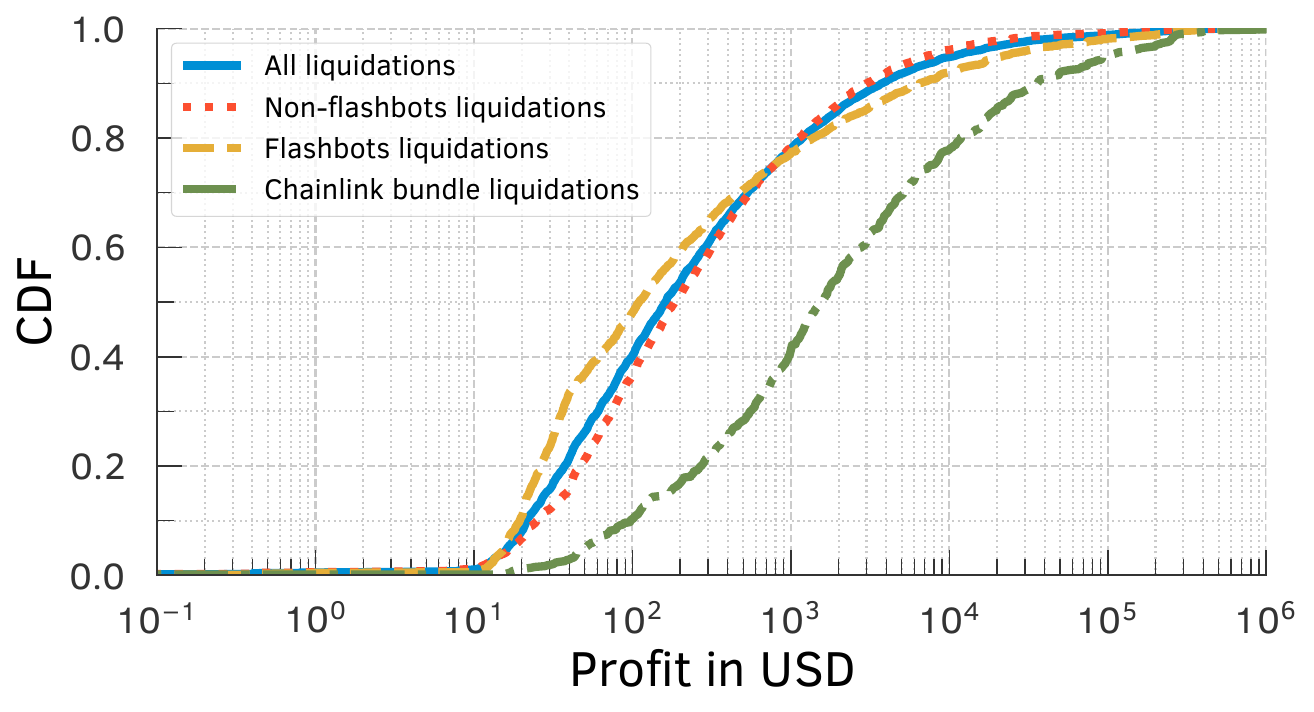}
    \sfigcap{Liquidations profit in Compound }\label{fig:oracle-update-Compound}
  \end{subfigure}
  \figcap{
  Profits of liquidators in (a) AAVE and in (b) Compound. Liquidations bundled with Chainlink updates generally provide higher profits.}
  \label{fig:oracle-update}
\end{figure*}

In order to calculate the profit made by the liquidators, we get the amount of debt that was repaid and the amount of the underlying collateral that was received by the liquidator. 
We calculate the price of each token at the time of liquidation by looking at the on-chain oracle price from Chainlink at the same block number, where the liquidation took place. 
For AAVE and Compound, we specifically use the Chainlink on-chain price used by AAVE and Compound in their respective protocols. 
AAVE uses the price in ETH as a reference for its tokens, whereas Compound's price oracles are denominated in USD. For AAVE, in order to calculate the profit made by each liquidation, we calculate the profit in ETH, and then multiply the profit by the current Chainlink on-chain price of ETH in USD.
Per Figure~\ref{fig:oracle-update}, liquidations that are bundled with a Chainlink update also have larger profits for liquidators, which implies that the lucrative liquidations are more likely to be bundled together with a Chainlink update.

\paraib{Characterizing transaction bundling}
To investigate which DEXes protocols are called within Flashbots bundles, we focus on the following contract calls: 0x Protocol~\cite{0xProtocol}, Balancer~\cite{Balancer}, Bancor~\cite{Bancor}, Curve~\cite{Curve}, SushiSwap~\cite{SushiSwap}, and Uniswap V1 and V3~\cite{Uniswap}.
In our set of \num{3284886} Flashbots bundles, we find that \num{2231051} (\num{67.92}\%) unique Flashbots bundles (and \num{3076760} transactions) called at least one of these contracts. Table~\ref{tab:flashbots-lending} shows the distribution of the number of transactions and the number of bundles for each of these contracts.
We see that Uniswap and SushiSwap are the most bundled DEXes protocols in Flashbots.

\begin{table}[]
\begin{center}
\tabcap{There are \num{2231051} (\num{67.92}\%) unique Flashbots bundles, and \num{3076760} (\num{44.35}\%) transactions, that called the following decentralized exchange contracts in Ethereum: 0x Protocol, Balancer, Bancor, Curve, SushiSwap, Uniswap V1, or V3. Note that a single transaction or bundle might call one or more contracts.}\label{tab:flashbots-lending}
\resizebox{\textwidth}{!}{%
\begin{tabular}{llllllll}
\toprule
\multicolumn{1}{c}{\thead{}} & \multicolumn{1}{c}{\thead{Balancer}}                          & \thead{Bancor}                                               & \multicolumn{1}{c}{\begin{tabular}[c]{@{}c@{}}\thead{Curve}\\ \thead{v1 \& v2}\end{tabular}} & \multicolumn{1}{c}{\begin{tabular}[c]{@{}c@{}}\thead{Uniswap v2}\\ \thead{\& Sushiswap}\end{tabular}} & \multicolumn{1}{c}{\begin{tabular}[c]{@{}c@{}}\thead{Uniswap}\\ \thead{v3}\end{tabular}} & \multicolumn{1}{c}{\begin{tabular}[c]{@{}c@{}}\thead{0x Protocol}\\ \thead{v1, v2 \& v3}\end{tabular}} & \thead{Total}                                                  \\ \midrule
\# of bundles                 & \begin{tabular}[c]{@{}l@{}}\num{85422}\\ \num{3.83}\%\end{tabular} & \begin{tabular}[c]{@{}l@{}}\num{96122}\\ \num{4.31}\%\end{tabular} & \begin{tabular}[c]{@{}l@{}}\num{53296}\\ \num{2.39}\%\end{tabular}                        & \begin{tabular}[c]{@{}l@{}}\num{1710985}\\ \num{76.69}\%\end{tabular}                                 & \begin{tabular}[c]{@{}l@{}}\num{1337715}\\ \num{59.96}\%\end{tabular}                    & \begin{tabular}[c]{@{}l@{}}\num{28753}\\ \num{1.29}\%\end{tabular}   & \begin{tabular}[c]{@{}l@{}}\num{2231051}\\ \num{67.92}\%\end{tabular} \\
\# of transactions                     & \begin{tabular}[c]{@{}l@{}}\num{87865}\\ \num{2.86}\%\end{tabular} & \begin{tabular}[c]{@{}l@{}}\num{99040}\\ \num{3.22}\%\end{tabular}  & \begin{tabular}[c]{@{}l@{}}\num{58188}\\ \num{1.89}\%\end{tabular}                         & \begin{tabular}[c]{@{}l@{}}\num{2533084}\\ \num{82.33}\%\end{tabular}                                & \begin{tabular}[c]{@{}l@{}}\num{1692485}\\ \num{55.01}\%\end{tabular}                   & \begin{tabular}[c]{@{}l@{}}\num{29100}\\ \num{0.95}\%\end{tabular}                                   & \begin{tabular}[c]{@{}l@{}}\num{3076760}\\ \num{44.35}\%\end{tabular} \\ \bottomrule  
\end{tabular}
}
  \end{center}
\end{table}

\subsection{Side channel (dark-fee) payments and transaction acceleration}

We now focus on the Bitcoin blockchain to study dark-fees transactions. 

\paraib{Prevalence of transaction acceleration}
Dark-fee transactions (or accelerated transactions) are transactions that offer additional fees to specific mining pools via an opaque and non-public side-channel payment~\cite{Messias@IMC2021}.
Messias~\ea show that in Bitcoin the top 5 mining pools, BTC.com~\cite{BTC@accelerator}, AntPool~\cite{AntPool@accelerator}, ViaBTC~\cite{ViaBTC@accelerator}, F2Pool~\cite{F2Pool@accelerator}, and Poolin~\cite{Poolin@accelerator}, deploy transaction acceleration services, which enables users to ``accelerate'' the confirmation of their transactions by offering mining pools dark-fees~\cite{Messias@IMC2021}.
These (dark-)fees are paid in fiat currency through a direct bank transfer or via other crypto coins to the mining pool. They are, therefore, opaque or dark to other participants.
Strangely enough, these fees are also non-refundable as the miner receives them regardless of whether they include the transaction in a block or not---a guaranteed payment.
The fees paid by the transaction issuer are, furthermore, not made public:
only the user and the miner knows the actual fee paid by the transaction inclusion.
Since transaction issuers pay the fees off-chain, miners have an incentive for prioritizing these transactions despite the low fee rate offered on-chain.
It also implies that the transaction issuer offers a miner a different fee compared to that offered to other miners for including their transaction in a block.
Miners do not disclose such private fees paid by issuers.
This behavior is different from that of Flashbots in Ethereum: The latter discloses the final dark-fee after the transaction is committed (see \S\ref{subsec:prevalence_of_flashbots}).

\begin{figure*}[tb]
  \centering
  \begin{subfigure}[b]{\onecolgrid}
    \includegraphics[width={\textwidth}]{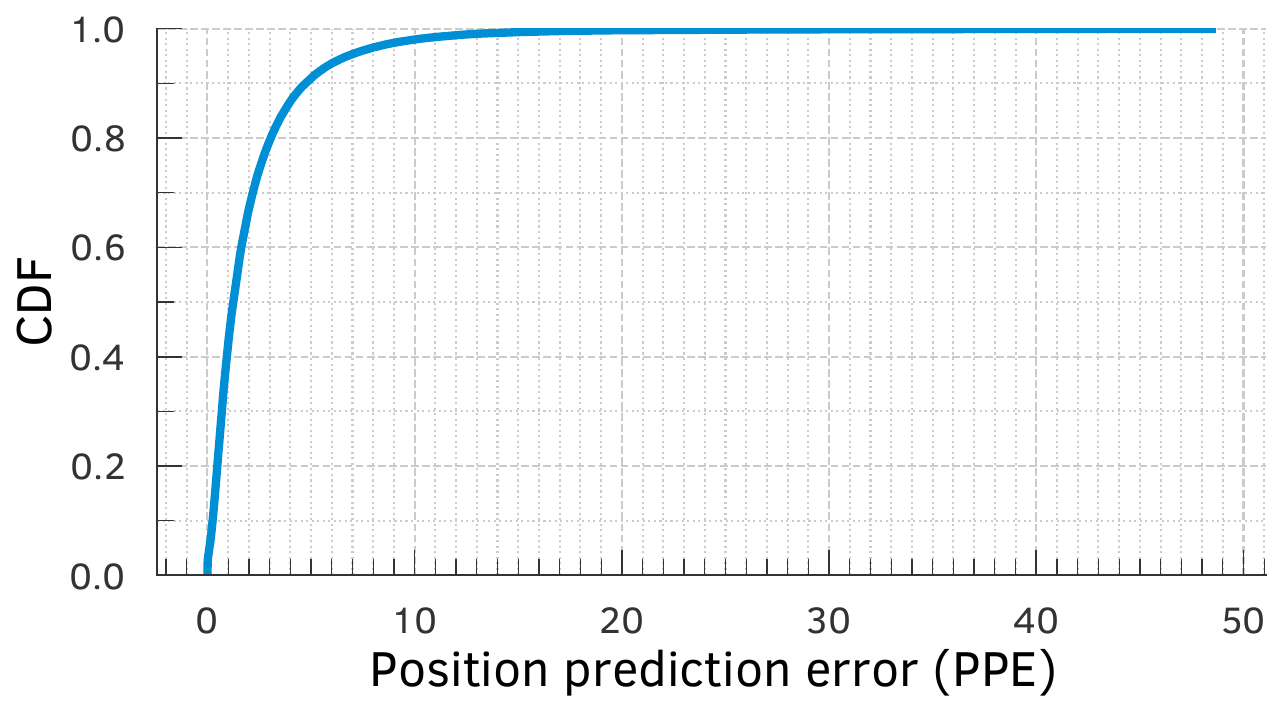}
    \figcap{Overall PPE}\label{fig:deviation-within-blocks-overall}
  \end{subfigure}
  \begin{subfigure}[b]{\onecolgrid}
    \includegraphics[width={\textwidth}]{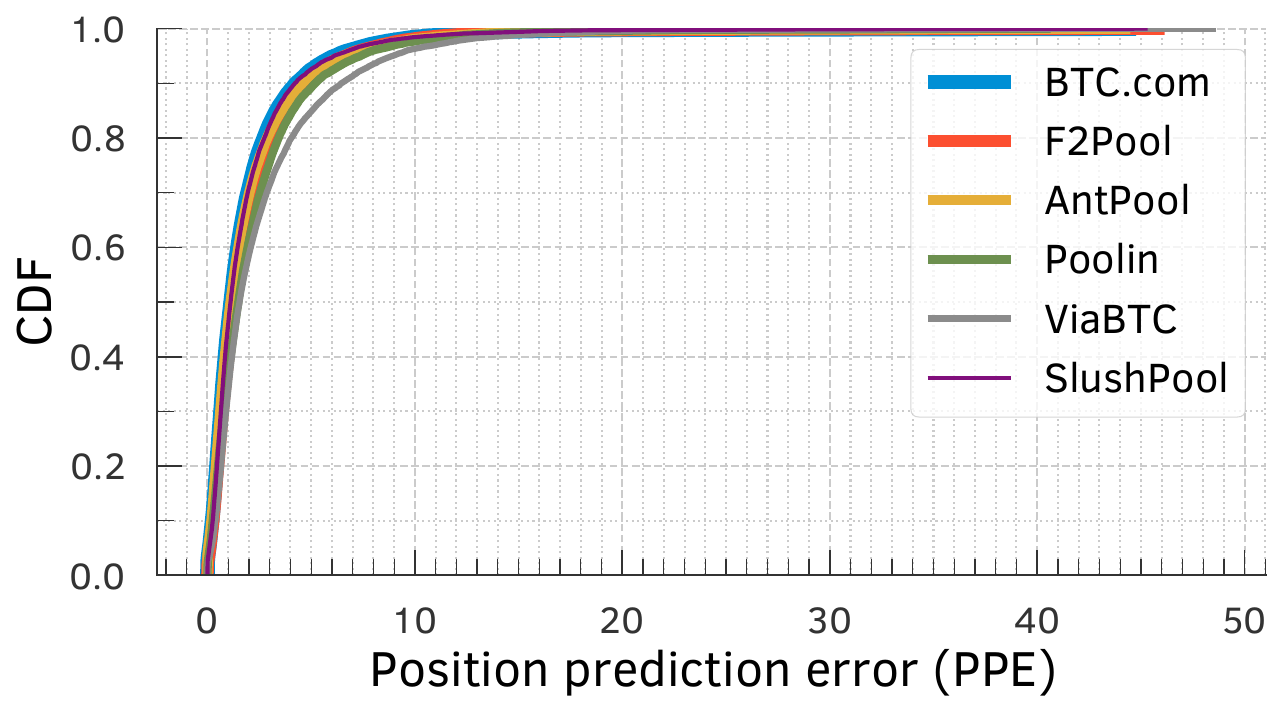}
    \figcap{PPE of the top-6 mining pools}\label{fig:deviation-within-blocks-top6-mpo}
  \end{subfigure}
  \figcap{
  Bitcoin position prediction error (PPE). (a) There are 160,962 blocks with non-CPFP txs; 80\% of all blocks has PPE less than 3.06\% (mean is 2.09\% and std. deviation is $2.75$.). (b) PPEs of top-6 mining pools per their normalized hash rate, showing that all large mining pools engage in transaction acceleration.
    }
  \label{fig:deviation-within-blocks}
\end{figure*}

\paraib{Characterizing transaction acceleration}\label{subsubsec:bitcoin-acceleration}
In order to detect accelerated transactions, Messias~\ea~\cite{Messias@IMC2021} proposed a metric called \stress{signed position prediction error (SPPE)} and \stress{position prediction error (PPE)}. 
The idea behind these measures is that transactions that have been accelerated through off-chain fees are likely to have been “misplaced” in a block based on the on-chain fee they offer.
Figure~\ref{fig:deviation-within-blocks} shows that the top-6 mining pools in our Bitcoin data set engage in transaction acceleration. 
Large SPPE values imply that a transaction that should have been included at the bottom is included at the top of the block, confirming acceleration. We rely on this methodology to infer transaction acceleration in Bitcoin and present our data set and findings below.

To identify accelerated transactions, we gathered all Bitcoin blocks mined from Jan. $1\tsup{st}$ 2018 to Dec. $31\tsup{st}~2020$. In total, there are \num{161954} blocks from block height \num{501951} to \num{663904}, and \num{313575387} transactions. In Bitcoin, mining pools may indicate their ownership of the block by including a \stress{signature} or \stress{marker} in the \stress{Coinbase}
transaction (i.e., the first transaction of every block).
We used such markers for identifying the mining pool (owner) of each block following techniques from prior work~\cite{judmayer2017merged,Messias@IMC2021,Romiti2019ADD}.
We failed to identify, however, the owners of $\num{4911}$ blocks (approximately $3\%$ of the blocks) and grouped these blocks under the label ``Unknown.''
Figure~\ref{fig:dist-tx-blks-bitcoin} shows the distribution of the count of blocks mined and transactions confirmed by the top-20 mining pools.
We further removed \num{65902514} (21.02\%) \stress{child-pays-for-parent (CPFP)} transactions from our acceleration analyses.

\begin{figure}[tb]
  \centering
    \includegraphics[width={\onecolgrid}]{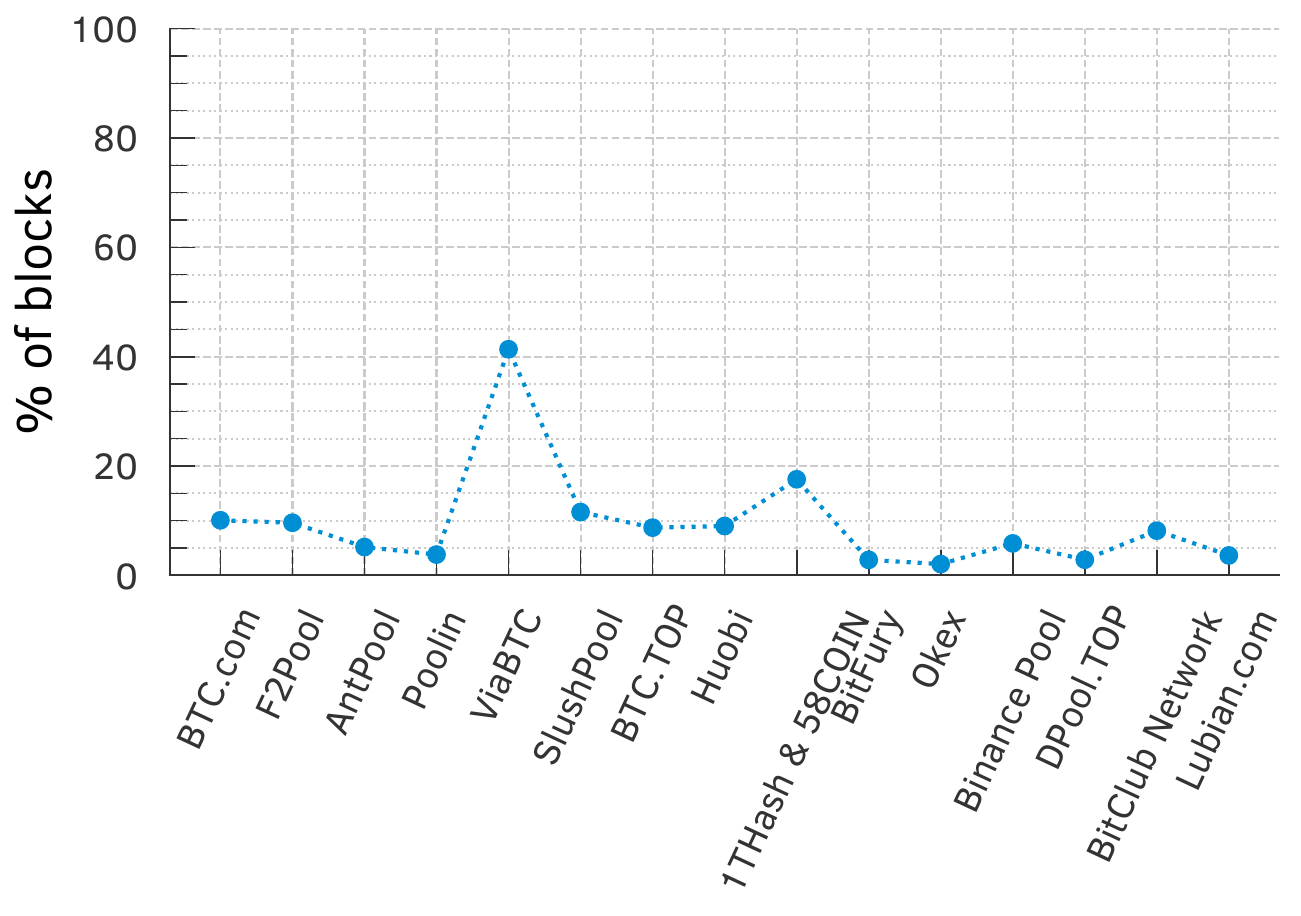}
  \figcap{
  Blocks with accelerated transactions (with SPPE $\geq 99\%$) are quite common among the top 15 mining pools. In Bitcoin, the mining pools with a high percentage of such blocks are ViaBTC (41.36\%), 1THash \& 58COIN (17.58\%), SlushPool (11.58\%), BTC.com (10.03\%), and F2Pool (9.63\%).}
  \label{fig:block-with-tx-violation}
\end{figure}

To estimate the prevalence of accelerated transactions in blocks mined by different mining pools, we compute the fraction of blocks mined by the top-15 mining pools, based on their hash rates in our data set (refer to \S\ref{sec:hash-var} and Figure~\ref{fig:dist-tx-blks-bitcoin}), that contained transactions with SPPE $\ge 99\%$. 
Per Figure~\ref{fig:block-with-tx-violation},
we find that many large mining pools such as BTC.com, F2Pool, and ViaBTC are likely including accelerated transactions in a sizeable fraction of their mined blocks, with ViaBTC including it in over $40\%$ of their blocks.

If we consider all mining pools' transactions with an SPPE $\geq 50\%$ (\num{1869043} transactions, in total), from $2018$ to $2020$, users transferred in total \num{11631217} BTC (or $\approx 223.55$ billion USD\footnote{Based on the Bitcoin exchange rate on October 19\tsup{th} 2022, 1 BTC = \num{19219.90} USD}). The accelerated transactions accounted for \num{240226} BTC (or $\approx 4.62$ billion USD), corresponding to approximately $2.07\%$. 

\paraib{Aggregated power of colluding miners}
In order to check the impact of transactions acceleration services on commit time of transaction, we ran active real-world experiments.
Specifically, we paid ViaBTC~\cite{ViaBTC@accelerator} to accelerate selected transactions (see Table~\ref{tab:acceleration-experiment} in \S\ref{sec:accelerated-txs}) during periods of high congestion between November 26\tsup{th} and December 1\tsup{st}, 2020.
From \num{10} \mpool snapshots during this period, we selected transactions that offered a very low fee-rate (i.e., 1--2 sat-per-byte) for acceleration.
To keep our acceleration costs low, we selected transactions with the smallest size (which was \num{110} bytes) within this set.
For each of the \num{10} snapshots, we had multiple transactions with such low fee-rates and small size, for a total of \num{212} transactions across all the snapshots.
We randomly selected one transaction from each snapshot (i.e., 10 transactions) and paid ViaBTC $205$ EUR to accelerate them.

\begin{table}[tb]
    \begin{center}
    \tabcap{Accelerated transactions have fewer delays and are included at the top of the block, i.e., at higher positions compared to non-accelerated transactions.}\label{tab:active-experiment-delay-position}
    \resizebox{.75\textwidth}{!}{%
        \begin{tabular}{rcccc}
        \toprule
        \multicolumn{1}{c}{\multirow{2}{*}{\thead{metrics}}} & \multicolumn{2}{c}{\thead{delay in \# of blocks}} & \multicolumn{2}{c}{\thead{perc. position in a block}} \\
        \multicolumn{1}{c}{}                         & \thead{acc.}        & \thead{non-acc.}       & \thead{acc.}       & \thead{non-acc.}      \\  \midrule
        minimum                                       & 1                  & 9                    & 0.07                 & 17.47                   \\
        25-perc                                       & 1                  & 148                    & 0.08                 & 75.88                   \\
        median                                        & 2                  & 191                    & 0.09                 & 87.92                   \\
        75-perc                                       & 2                  & 247                    & 0.20              & 95.00                   \\
        maximum                                       & 3                  & 326                    & 4.39               & 99.95                   \\
        average                                       & 1.8                & 198.5                    & 0.79              & 84.46                   \\
        \bottomrule
        \end{tabular}
    } 
    \end{center}
\end{table}

We then compare the priority with which the accelerated transactions and the $202$ ($= 212-10$) non-accelerated transactions with similar fee rates and sizes were included in the Bitcoin blockchain.
The impact of acceleration was strikingly apparent as shown in Table~\ref{tab:active-experiment-delay-position}. 
All $10$ accelerated transactions were included within $1$--$3$ blocks after their acceleration, with an average delay of $1.8$ blocks.
In contrast, the minimum delay for the $202$ non-accelerated transactions of comparable fee-rates and sizes was $9$ blocks, with an average delay of 198.5 blocks.
Interestingly, $38$ of the non-accelerated transactions were yet to be included in the blockchain by December 4\tsup{th}, 2020.
Similarly, the accelerated transactions were included in top $0.07$--$4.39$ percentile positions, with an average $0.79$ percentile position, while the non-accelerated transactions were included in the beyond top $17.47$--$99.95$ percentile positions, with an average $84.46$ percentile position.
From the above observations, it is clear that the transactions we accelerated were included with high priority, meaning Bitcoin mining pools take off-chain fees into account when prioritizing transactions.

Although, we accelerated our transactions using ViaBTC mining pool, our $10$ transactions were included by $5$ different mining pools, namely F2Pool, AntPool, Binance, Huobi, and ViaBTC. 
As we accelerated transaction during time of high congestion in Bitcoin, no mining pool would have included a transaction offering $1$--$2$ sat-per-byte, unless they were accelerated. Since we only paid the ViaBTC mining pool, this implies that ViaBTC is colluding with other mining pools to accelerate transactions that offer off-chain fees.
Except for Binance, all these colluding pools rank amongst the top-$8$ mining pools in terms of their hash rates at the time of our experiments.
Table~\ref{tab:active-experiment-hash-rate} shows the individual as well as the combined hash rates of these $5$ colluding mining pools over the last day, last week, and last month before the conclusion of our experiment on December $1\tsup{st}, 2020$. 
The most striking and the most worrisome fact is that the combined hash rates of these colluding mining pools exceeds $55\%$ of the total Bitcoin hash rate. For more details, refer to Figures~\ref{fig:tx-acceleration-active-overtime-month} and \ref{fig:tx-acceleration-passive-active-overtime-month} in \S\ref{sec:accelerated-txs} in the appendix. Additionally, if mining pools are colluding to include accelerated transactions, then they might also potentially collude in malicious ways.  

\begin{table}[tb]
    \begin{center}
    \tabcap{If we rank the miners who confirmed the accelerated transactions based on their daily, weekly, and monthly hash rate power, at the time these experiments were conducted, the combined hash power of these mining pools exceeds 55\% of the Bitcoin's total hashing power.}\label{tab:active-experiment-hash-rate}
    \resizebox{.65\textwidth}{!}{%
        \begin{tabular}{rccc}
        \toprule
        \multicolumn{1}{c}{\multirow{2}{*}{\thead{Mining Pool}}} & \multicolumn{3}{c}{\thead{Hash-rate}}                                                                                \\
        \multicolumn{1}{c}{}                     & \multicolumn{1}{c}{\thead{last 24h}} & \multicolumn{1}{c}{\thead{last week}} & \multicolumn{1}{c}{\thead{last month}} \\ \midrule
        F2Pool                                    & \multicolumn{1}{c}{$19.9\%$}   & \multicolumn{1}{c}{$18.7\%$}    & 
        \multicolumn{1}{c}{$19.9\%$}    \\
        AntPool                                   & \multicolumn{1}{c}{$12.5\%$}   & \multicolumn{1}{c}{$10.6\%$}    & \multicolumn{1}{c}{$10.2\%$}    \\
        Binance                                   & \multicolumn{1}{c}{$9.6\%$}    & \multicolumn{1}{c}{$10.3\%$}    & \multicolumn{1}{c}{$10.0\%$}    \\
        Huobi                                     & \multicolumn{1}{c}{$8.1\%$}    & \multicolumn{1}{c}{$9.3\%$}     & \multicolumn{1}{c}{$9.8\%$}     \\
        ViaBTC                                    & \multicolumn{1}{c}{$5.1\%$}    & \multicolumn{1}{c}{$7.1\%$}     & \multicolumn{1}{c}{$7.7\%$}     \\
        \thead{Total}                                     & \multicolumn{1}{c}{\attention{$55.2\%$}}   & \multicolumn{1}{c}{\attention{$56\%$}}      & \multicolumn{1}{c}{\attention{$57.6\%$}}   \\ \bottomrule
        \end{tabular}
    } 
    \end{center}
\end{table}

Furthermore, due to the lack of transparency into their queue, miners can charge higher prices for their acceleration services when colluding. It means that they can overcharge the transaction issuers for including their transactions.

\section{Concluding Discussion}\label{sec:conclusion}

In this section, we discuss the implications of our findings regarding the lack of transparency in transaction contention and prioritization.
We also argue why our findings and implications would be relevant even in the face of recent changes to blockchain protocols, e.g., Ethereum Improvement Protocol (EIP) 1559 and the Ethereum Paris Network Upgrade (a.k.a. the Merge).

\parai{Implications for publicly mined transactions}
Most wallet software and crypto-exchanges today rely on reconstructing the current public \mpool state in order to suggest a suitable fee to transaction issuers~\cite{Wallet@BitcoinCore,Fees@Coinbase,Wallet@Electrum,Wallet@Metamask,Wallet@Trezor}. 
With the lack of contention and prioritization transparency, transaction issuers can no longer accurately recreate the current \mpool state for different miners. 
Consequently, they cannot reliably estimate the fees transactions need to pay for their desired prioritization. 
Worse, as the fraction of privately mined and accelerated transactions keeps rising, the transaction fees will become less (reliably) predictable in the future. 

\parai{Implications for privately mined transactions}
The problem of reliable fee estimation for a desired level of prioritization is even worse for privately mined transactions that are announced on private relay networks.
When transaction issuers announce on a private relay network today, they are often unsure what fraction of total hash rate is controlled by the miners listening to the private relay network.
It is important to estimate the hash rate controlled by private mining pools to estimate the commit (waiting) times for transactions.
Furthermore, transaction issuers on private relay networks are completely blind to other competing transactions.
This opacity allows miners offering private mining and transaction acceleration services to overcharge and demand exorbitant fees to commit transactions.
For example, in the Ethereum blockchain, users are observed to be overcharged by miners for having their transactions confirmed with high priority through Flashbots bundles~\cite{Weintraub@IMC2022}.

\parai{Relevance of findings in light of EIP-1559 and the Ethereum Merge} 
Our observations about the lack of transparency and their implications are fundamental to the current blockchain architectures and hold both before and after the recent major improvements to blockchains, e.g., EIP-1559 and the Ethereum Merge.
While EIP-1559 attempts to improve the estimation of transaction fees that need to be offered, it does not address the problems associated with the lack of transaction contention and prioritization transparency. 
Similarly, after the Ethereum Merge, \stress{validators} that stake a certain amount of ETH rather than \stress{miners} would be responsible for selecting and validating transactions to include in the next block~\cite{Eth-PoS}.
Our observations about private mining would still hold for private validation and the implications would still be valid after the Merge.

In conclusion, our work shows that with private mining and accelerated transactions, the promise of the public decentralized blockchain does not hold. 
Firstly, mining pools with combined hash rates of over $50\%$ are colluding with each other, showing a centralization in the system.
Then, they can also censor certain transactions, breaking the ethos of decentralized public blockchains with no central authorities. 
Second, it breaks the assumption that all activities in the blockchain are transparent.
Although this is true for transactions included in the blockchain, prioritization of transactions is becoming more opaque with the rise of private mining and off-chain fees.
Hence, we make the case that to fulfill the transparency promise of public blockchains, prioritization of transactions should be transparent as well.
Third, with private mining in Ethereum, Flashbots is increasingly being used for malicious and predatory activities such as sandwich attacks, which essentially levies a tax on users interacting with financial institutions on the blockchain (e.g., in DEX).
These concerns need to be addressed if public blockchains are going to live up to their promises. 

\section*{Acknowledgments}\label{sec:ack}

This research was supported in part by a European Research Council (ERC) Advanced Grant ``Foundations for Fair Social Computing'', funded under the European Union's Horizon 2020 Framework Programme (grant agreement no. 789373). It was also supported by MIAI @ Grenoble Alpes (ANR-19-P3IA-0003) and by the French National Research Agency under grant ANR-20-CE23-0007.

\bibliographystyle{splncs04}
\bibliography{references}

\appendix

\section{Ethereum private transaction experiment} \label{sec:private-txs}

We conducted $4$ active experiments where we issued $8$ Ethereum transactions; half issued publicly and the other half privately through a private-channel network known as Taichi Network~\cite{Taichi@accelerator}. Table~\ref{tab:acceleration-experiment-ETH} summarizes the transactions in our experiment. Spark Pool and Babel Pool included all private transactions ($2$ transactions each) sent directly to these miners through Taichi Network.

\begin{table*}[]
\caption{We conducted 4 active experiments in Ethereum by simultaneously accelerating transactions privately and publicly via Taichi Network. Private transactions were included only by Spark Pool and Babel Pool. If we rank these mining pools according to their hash-rate, they account for 27.72\% of the total Ethereum hash-rate.}
\label{tab:acceleration-experiment-ETH}
\resizebox{\textwidth}{!}{%
\begin{tabular}{rrrcrrccrccrc}
\hline
\multirow{2}{*}{\thead{\#}} & \multirow{2}{*}{\thead{type}} & \multirow{2}{*}{\thead{tx hash}}                                           & \multirow{2}{*}{\thead{block number}} & \multirow{2}{*}{\thead{miner}} & \thead{tx. position}   & \thead{block delay} & \thead{fee paid}                   & \thead{base fee} & \thead{max fee} & \thead{max priority fee} & \thead{gas price}     & \thead{block timestamp}     \\  
                    &                       &                                                                    &                               &                        & \thead{per \# of txs.} & \thead{(in blocks)} & \thead{(in Ether)}                 & \thead{(Gwei)}           & \thead{(Gwei)}          & \thead{(Gwei)}                   & \thead{(Gwei)}        & \thead{in UTC}              \\ \hline
\multirow{2}{*}{1}  & public                & \href{https://etherscan.io/tx/0xbbe88eae757acf6697d498575dd1d50b3ad9915318cd1ff8d409210d20a4f000}{bbe88e$\cdots$a4f000} & 13,183,516                    & Nanopool               & 305 / 336      & 1           & 0.00190489 & 88.98082939     & 116.52835749   & 1.72836605              & 90.70919543  & 2021-09-08 06:39:18 \\
                    & \thead{private}               & \href{https://etherscan.io/tx/0xc46b7556a20865c9f50166373baf7094104f300ab26ad8e1de894e1318ead538}{c46b75$\cdots$ead538} & 13,183,520                    & Babel Pool             & 29 / 39        & 5           & 0.00225209  & 105.51391459    & 120.56586232   & 1.72836605              & 107.24228063  & 2021-09-08 06:40:29 \\
\multirow{2}{*}{2}  & public                & \href{https://etherscan.io/tx/0x6d994f516f43b8ed3763fe4f81c7cb86146203fda1047cc85e697eefa7c1aadd}{6d994f$\cdots$c1aadd} & 13,183,561                    & Binance                & 209 / 213      & 2           & 0.00244137 & 114.95482846    & 137.64014705   & 1.30100683              & 116.25583529 & 2021-09-08 06:49:26 \\
                    & \thead{private}               & \href{https://etherscan.io/tx/0xa4d4ae2f6f3a798dc6cf5d5f4e15222320d3ee90b023763efe0017e51142ebf5}{a4d4ae$\cdots$42ebf5} & 13,183,565                    & Spark Pool             & 294 / 296      & 6           & 0.00240978 & 113.45059961    & 137.64014705   & 1.30100683              & 114.75160643 & 2021-09-08 06:50:12 \\
\multirow{2}{*}{3}  & public                & \href{https://etherscan.io/tx/0x725743c1700241a6e89b957faf963018f2d169f7f1ec6b9256a92811510a6c45}{725743$\cdots$0a6c45} & 13,183,634                    & Unknown                & 124 / 126      & 2           & 0.00263298 & 123.27216185    & 135.21393222   & 2.10805685              & 125.38021870 & 2021-09-08 07:06:31 \\
                    & \thead{private}               & \href{https://etherscan.io/tx/0xf2beec913ed6c0667fdde4829a004fe9418916af22218d77adf5f38a7c15cdf1}{f2beec$\cdots$15cdf1} & 13,183,635                    & Spark Pool             & 321 / 340      & 3           & 0.00257468 & 120.49562077     & 135.21393222   & 2.10805685              & 122.60367762 & 2021-09-08 07:06:44 \\
\multirow{2}{*}{4}  & public                & \href{https://etherscan.io/tx/0xe21695cc9e1f29f45f38b0fd8323a6e928bd7b55dc84974f217c7042322c1574}{e21695$\cdots$2c1574} & 13,183,679                    & Ethermine              & 280 / 302      & 13          & 0.00223433 & 104.69510748    & 108.95262574   & 1.70164453              & 106.39675202 & 2021-09-08 07:18:37 \\
                    & \thead{private}               & \href{https://etherscan.io/tx/0x4c482b0416b38de9b2995b986d8c0f974018c0aeda02ce6fdc8b196bce87c76f}{4c482b$\cdots$87c76f} & 13,183,690                      & Babel Pool             & 150 / 212      & 24          & 0.00179917  & 83.97323655     & 108.95262574   & 1.70164453              & 85.67488108   & 2021-09-08 07:20:12 \\ \hline
\end{tabular}%
}
\end{table*}

\section{Liquidation with Chainlink oracle updates}\label{sec:liquidations-cll-updates}

In AAVE, of \num{1154} bundles, \num{994} (86.14\%) include one Chainlink oracle update followed by a liquidation. There are \num{52} (4.51\%) with two oracle updates followed by liquidations.
Out of \num{1301} oracle updates bundled with liquidations, \num{282} ($21.68\%$) are USDC-ETH, \num{203} ($15.60\%$) are USDT-ETH, \num{169} ($12.99\%$) are DAI-ETH, \num{70} ($5.38\%$) are SUSD-ETH, and \num{60} ($4.61\%$) are LINK-ETH.
In Compound, of \num{641} bundles, \num{548} (85.49\%) contain one Chainlink oracle update followed by one liquidation, while \num{39} (6.08\%) include two oracle updates followed by liquidations.
Out of \num{751} oracle updates bundled with liquidations, \num{311} ($41.41\%$) are ETH-USD, \num{128} ($17.04\%$) are BTC-USD, and \num{53} ($7.06\%$) are UNI-USD.

\section{Hashing rates of mining pools}\label{sec:hash-var}

\begin{figure*}[tb]
    \centering
    \includegraphics[width=\textwidth]{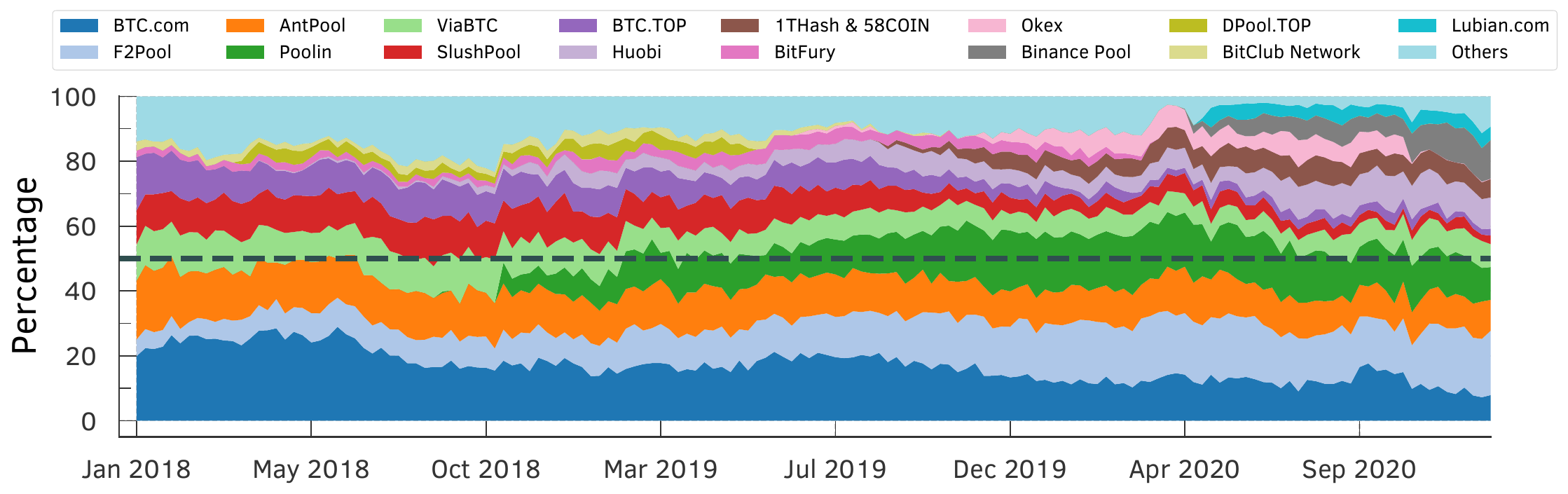}
    \figcap{Monthly Bitcoin hash rate over the 3-year period.}\label{fig:btc-hashrate}
\end{figure*}

Per Figure~\ref{fig:btc-hashrate}, the hash rates of Bitcoin mining pools such as BTC.com, F2Pool, and AntPool alone accounted for almost half the total hash rate of the network around May 2018, and roughly a year later, i.e., from March 2019, together with Poolin the four mining pools alone represent more than $50\%$ of the total network hash rate.
At the end of 2020, new MPOs, e.g., Lubian.com and Binance Pool, started mining Bitcoin, which help improve the decentralization of Bitcoin.
However, BTC.com, F2Pool, AntPool, and Poolin still account for almost half of the hash rates showing that a few mining pools control a considerable portion of the Bitcoin hash rate.

\begin{figure*}[tb]
    \centering
    \includegraphics[width=\textwidth]{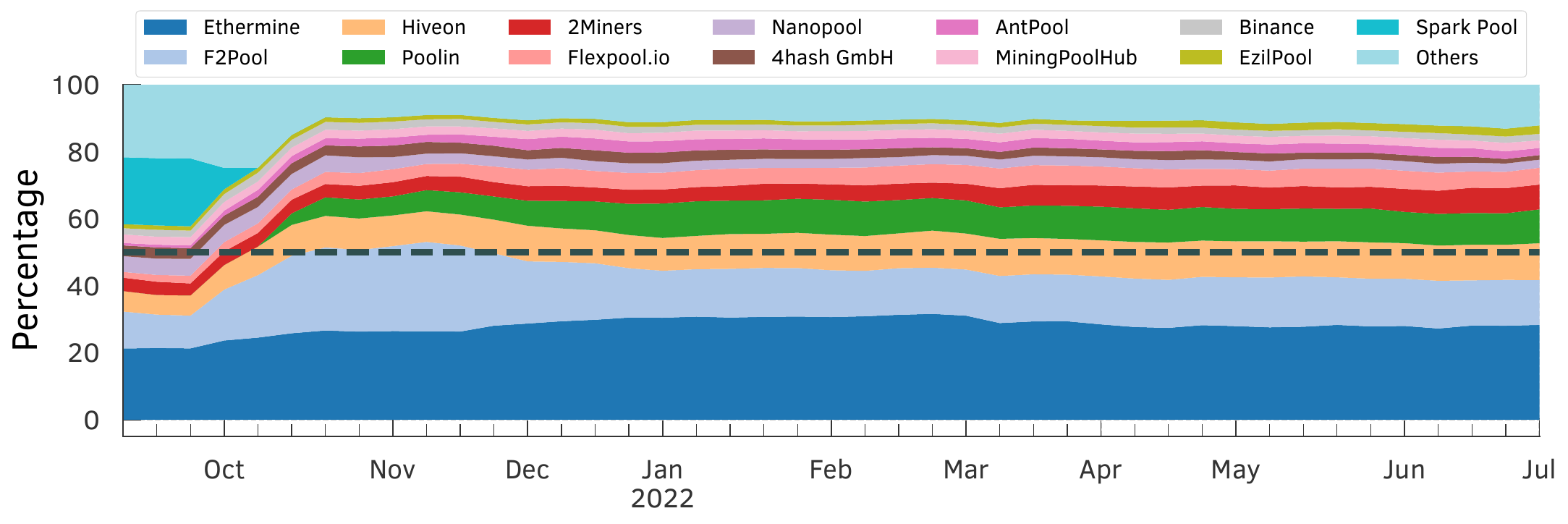}
    \figcap{Weekly Ethereum hash rate from Sept $8\tsup{th}$, 2021, to Jun $30\tsup{th}$, 2022.}\label{fig:eth-hashrate}
\end{figure*}

Hash rates of Ethereum mining pools, in contrast to Bitcoin, do \stress{not} show a high variance (see Figure~\ref{fig:eth-hashrate}).
We also observed that Spark Pool, the second-largest Ethereum mining pool, suspended their mining services on September 30, 2021, due to regulatory requirements in response to Chinese authorities~\cite{SparkPool@CoinTelegraph}.

\section{Bitcoin transaction acceleration experiment} \label{sec:accelerated-txs}

\begin{table*}[]
\caption{We conduct 10 transaction acceleration experiments in Bitcoin. If we rank the miners whose included these transactions based on their daily hash-rate power as (D) and weekly hash-rate power as (W), together these mining pools corresponds to a hash-rate power of (D: 55.2\%; W: 56\%).}
\label{tab:acceleration-experiment}
\resizebox{\textwidth}{!}{%
\begin{tabular}{rcccccccccc}
\toprule
\multicolumn{1}{c}{\multirow{2}{*}{\thead{txid}}}                       & \multirow{2}{*}{\thead{block height}} & \multirow{2}{*}{\thead{miner}} & \multirow{2}{*}{\thead{tx. position}} & \thead{delay}       & \thead{acc. cost} & \thead{vsize}  & \thead{fee rate}       & \multicolumn{2}{c}{\thead{Mempool}} & \thead{timestamp}        \\ 
\multicolumn{1}{c}{}                                            &                               &                        &                               & \thead{(in blocks)} & \thead{(BTC)}     & \thead{(byte)} & \thead{sat-per-vsize} & \thead{\# of txs.}    & \thead{vsize (MB)}    & \thead{in UTC}           \\ \midrule
\href{https://explorer.btc.com/btc/transaction/35b18e7a119173c8136c460e45d5d2a87d69304f69546f22ebed2c5f3852dbc1}{35b18e$\cdots$52dbc1} & \num{658805}                        & Huobi                  & \nth{2}                           & 2           & 0.001254  & 110    & 2             & \num{36644}        & 44.63   & 2020-11-26 19:10 \\
\href{https://explorer.btc.com/btc/transaction/65765c65acc86bde3d305b2594229af0839b3636aabea49e7255521412baede2}{65765c$\cdots$baede2} & \num{658898}                        & F2Pool                 & \nth{73}                          & 1           & 0.001254  & 110    & 2             & \num{20998}        & 32.55   & 2020-11-27 11:06 \\
\href{https://explorer.btc.com/btc/transaction/0c2098e3b3c993f5fc1d188da3b9d0a8731961bb946c4048d7a99fa83129fbf0}{0c2098$\cdots$29fbf0} & \num{658912}                        & AntPool                & \nth{2}                           & 2           & 0.001254  & 110    & 1             & \num{30126}        & 38.01   & 2020-11-27 13:38 \\
\href{https://explorer.btc.com/btc/transaction/1515a78b711558a1508400b36f554d798a31bd97e3852de5bae598e020179af3}{1515a7$\cdots$179af3} & \num{658971}                        & Binance                & \nth{2}                           & 3           & 0.001254  & 110    & 1             & \num{25922}        & 37.89   & 2020-11-27 21:55 \\
\href{https://explorer.btc.com/btc/transaction/48a0a55252bc029286e4af6215d1673e6744216ffc86b3c7b36eeafe640ddaec}{48a0a5$\cdots$0ddaec} & \num{659335}                        & ViaBTC                 & \nth{3}                           & 1           & 0.001045  & 110    & 1             & \num{15605}        & 9.82    & 2020-11-30 10:09 \\
\href{https://explorer.btc.com/btc/transaction/9a17cfef7e7bda668415a4a4918195669086f0507786a0c971df24a1c3f3734c}{9a17cf$\cdots$f3734c} & \num{659341}                        & Huobi                  & \nth{2}                           & 2           & 0.001045  & 110    & 1             & \num{14945}        & 9.41    & 2020-11-30 10:28 \\
\href{https://explorer.btc.com/btc/transaction/831b246f748db46d4f52318e39171b0b587165282be3f07135d978ef0795d421}{831b24$\cdots$95d421} & \num{659351}                        & AntPool                & \nth{2}                           & 1           & 0.001045  & 110    & 1             & \num{10990}        & 8.66    & 2020-11-30 12:22 \\
\href{https://explorer.btc.com/btc/transaction/1f59bfc1ef2de7b2bc9d3dd3f3e35dba437c25a93d53533a76d604284047096c}{1f59bf$\cdots$47096c} & \num{659355}                        & F2Pool                 & \nth{111}                         & 3           & 0.001045  & 110    & 1             & \num{17093}        & 11.40   & 2020-11-30 12:58 \\
\href{https://explorer.btc.com/btc/transaction/6942e0751586aa8f37b6cad4eb036373035d74f40ba36277a7d1ef17ca8c06c3}{6942e0$\cdots$8c06c3} & \num{659362}                        & Huobi                  & \nth{2}                           & 2           & 0.001045  & 110    & 1             & \num{30836}        & 19.06   & 2020-11-30 14:49 \\
\href{https://explorer.btc.com/btc/transaction/8e49e27c5eb6959e26dec8ab36d4dc6508105447ce8892d71c2837934eae825f}{8e49e2$\cdots$ae825f} & \num{659481}                        & ViaBTC                 & \nth{6}                           & 1           & 0.001254  & 110    & 2             & \num{30935}        & 22.59   & 2020-12-01 10:40 \\ \bottomrule
\end{tabular}
}
\end{table*}

\begin{figure*}[tb]
    \centering
    \includegraphics[width=\textwidth]{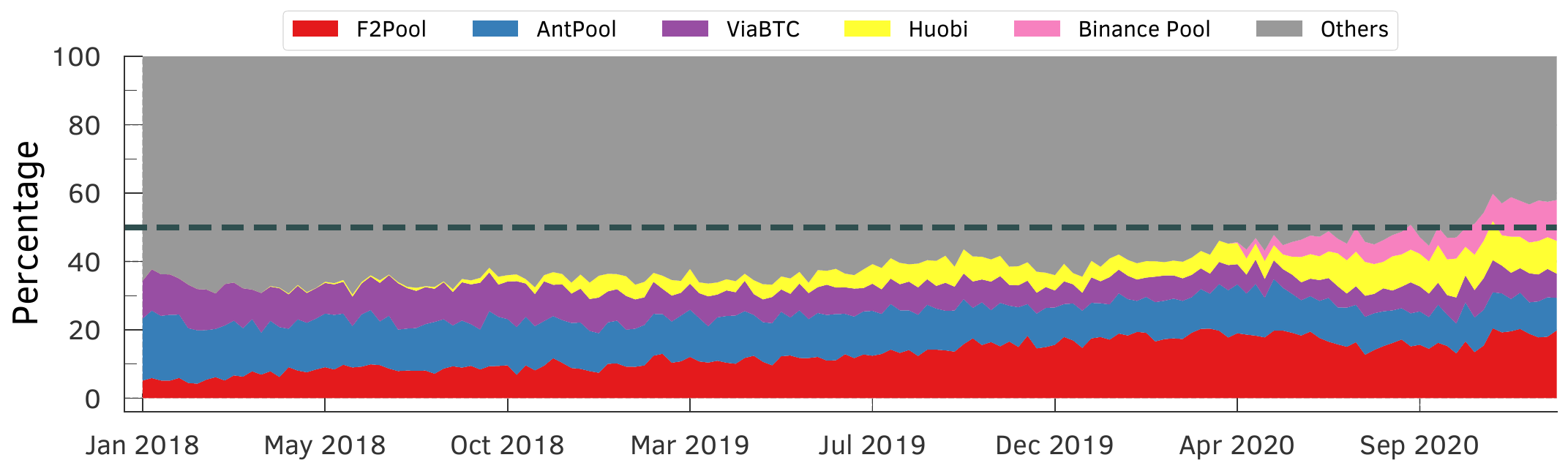}
    \figcap{Active vs. others experiment: Bitcoin mining pools in the active experiment (i.e., mining pools that included transactions accelerated by ourselves) increased their hash rate in 2020. Together, they accounted for more than $55\%$ of the overall hash rate. The plot shows the weekly average percentage of the mining pool's hash-rate over 3 years.}\label{fig:tx-acceleration-active-overtime-month}
\end{figure*}

\begin{figure*}[tb]
    \centering
    \includegraphics[width=\textwidth]{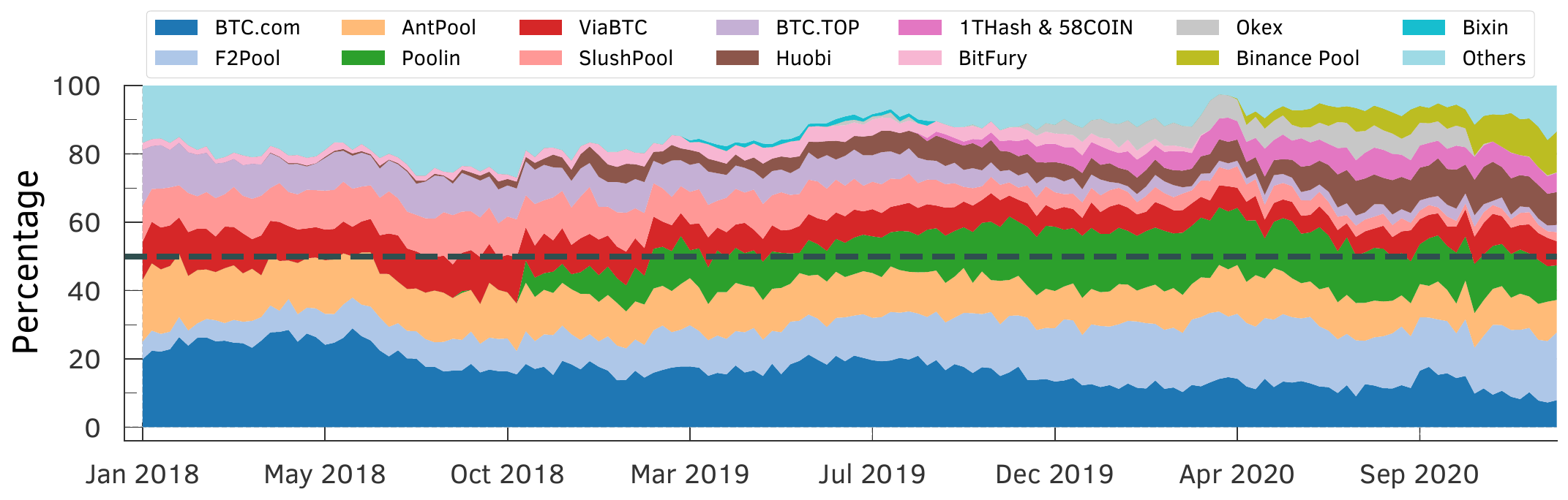}
    \figcap{Passive + active vs. others experiment: Bitcoin mining pools in the active experiment (i.e., mining pools that included transactions accelerated by ourselves) and passive experiment (mining pools that included transactions inferred to be accelerated using the BTC.com API) increased their hash rate in 2020. The plot shows the weekly average percentage of the mining pool's hash-rate over 3 years.}\label{fig:tx-acceleration-passive-active-overtime-month}
\end{figure*}

We ran an active Bitcoin transaction acceleration experiment where we paid $205$ EUR to ViaBTC~\cite{ViaBTC@accelerator} to accelerated $10$ transactions from $10$ different snapshots of our \mpool.
To select these transactions, we checked whether the \mpool was congested (i.e., having more transactions waiting for inclusion than the
next block would be able to include), with its size being at least $\uMB{8}$.
Then, we
considered only transactions with low fee rates---less than or equal to
2 sat-per-byte---to ensure that these transactions would be highly unlikely to be included soon in a subsequent block.
Next, we sorted the remaining transactions by size to limit the experiment cost
as the acceleration-service costs grow proportional to the transaction size. Finally, 
we select the transaction with the smallest size in bytes for our active experiment.

Most of these $10$ accelerated transactions were included nearly in the next block, demonstrating the acceleration efficiency. Also, these transactions were wrongly positioned in the block: They appeared, for instance, at the top of the block, i.e., higher than the non-accelerated transactions, showing that miners indeed prioritized them (see Table~\ref{tab:active-experiment-delay-position}).
Further, we observed that although we had only accelerated transactions via ViaBTC, other top mining pools were also involved in confirming the accelerated transactions.

Table~\ref{tab:acceleration-experiment} shows the transactions used in our experiments. At the time we conducted our experiments, if we rank the miners whose included these transactions based on their daily hash-rate power as (D) and weekly hash-rate power as (W), we would have Huobi (D: $8.1\%$; W: $9.3\%$), Binance (D: $9.6\%$; W: $10.3\%$), F2Pool (D: $19.9\%$; W: $18.7\%$), AntPool (D: $12.5\%$; W: $10.6\%$), ViaBTC (D: $5.1\%$; W: $7.1\%$). Together these mining pools corresponds to a hash-rate power of (D: $55.2\%$; W: $56\%$). Figures~\ref{fig:tx-acceleration-active-overtime-month} and \ref{fig:tx-acceleration-passive-active-overtime-month} show the hash-rate of mining pools in the active experiment and considering the passive experiment (inferred to be accelerated by BTC.com API), respectively.

\begin{figure*}[tb]
    \centering
    \includegraphics[width=\textwidth]{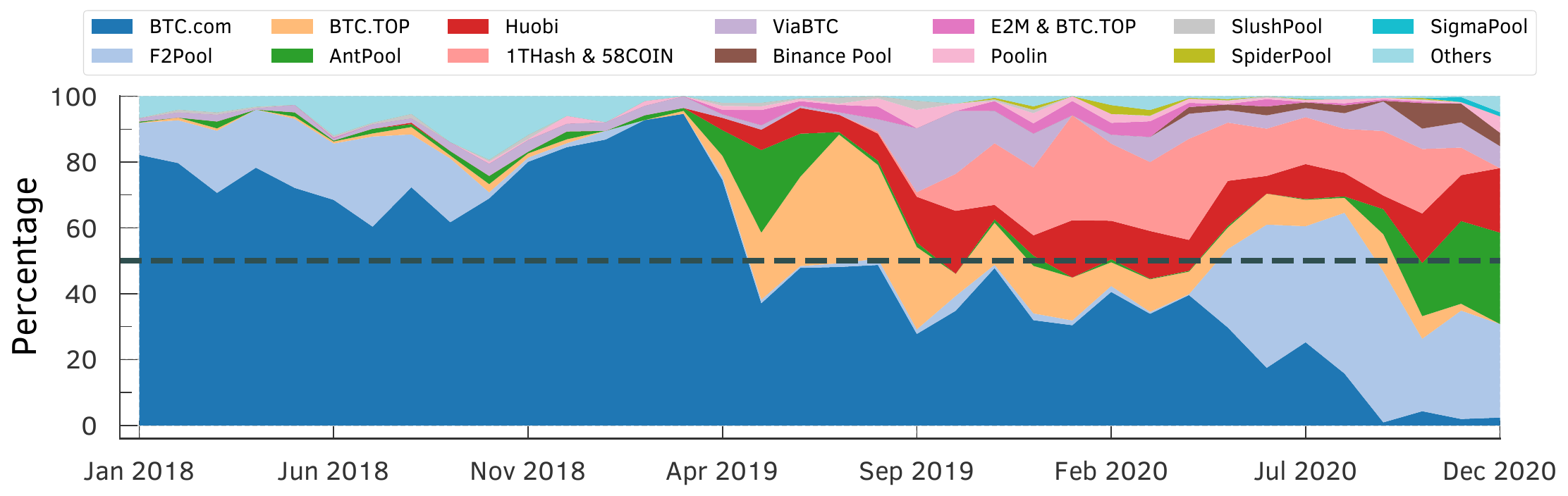}
    \figcap{The plot shows the monthly average percentage of accelerated Bitcoin transactions inclusion by each mining pool over 3 years. Transaction acceleration services or simply Front-running as a Services (FRaaS) are becoming popular across all mining pools.}\label{fig:tx-acceleration-overtime-month}
\end{figure*}

Furthermore, BTC.com~\cite{BTC@accelerator}, one of the leading Bitcoin mining pools, provides transaction acceleration services and allows users to verify if transactions have been accelerated through their platform or partner services.
From our dataset, we selected those with a SPPE greater than or equal to \num{1}\% (\num{12983282} transactions in total) and checked if they were said to be accelerated by BTC.com's API.
Of these transactions, \num{14104} were found to have been accelerated.
Our findings also show that transaction acceleration services are becoming quite common among Bitcoin mining pools (as shown in Figure~\ref{fig:tx-acceleration-overtime-month}).
Between 2018 and April 2019, only BTC.com and F2Pool alone accounted for most of the accelerated transactions.
However, as of December 2020, we see that BTC.com accounts for a very small fraction of accelerated transactions, with AntPool, Huobi, and F2Pool accounting for most of the accelerated transactions. 

\end{document}